\documentclass[aps,preprint,onecolumn,eqsecnum,nofootinbib]{revtex4}
\usepackage{dcolumn}
\usepackage{bm}
\usepackage{amssymb}
\usepackage{amsmath}
\usepackage{amsfonts}

\usepackage[]{graphicx}
\begin{document}
\bibliographystyle{prsty}
\title{ Arrival time and  Bohmian Mechanics: It is the theory which decides what we can measure}
\author{ Aur\'elien Drezet $^{1}$}
\address{(1) Univ. Grenoble Alpes, CNRS, Institut N\'{e}el, F-38000 Grenoble, France
}
\begin{abstract}
In this work we analyze recent proposals by Das and D\"{u}rr (DD) to measure the arrival time distributions of quantum particles within the framework of de Broglie Bohm theory (or Bohmian mechanics). We also analyze the criticisms made by Goldstein Tumulka and Zangh\`{i} (GTZ) of these same proposals, and show that each protagonist is both right and wrong. In fine, we show that DD's predictions are indeed measurable in principle, but that they will not lead to violations of the no-signalling theorem used in Bell's theorem, in contradiction with some of Das and Maudlin's hopes. 
\end{abstract}

\pacs{xx} \maketitle

\section{Introduction}
\label{section1}
The concept of arrival time for a quantum particle at a spatial point has been a subject of considerable controversy since the theory was founded in the 1920s-30s \cite{Muga2008} (Similar difficulties arise in defining travel and dwell times in quantum mechanics \cite{Field2022,Leavens1990}).  At the more technical level the main difficulty stems from the lack of consensus on the definition of a self-adjoint operator or POVM (positive operator valued measure) for the  arrival time $\tau$ of a particle, and on the probability distribution $\mathcal{P}^\Psi(\tau)$ associated with these arrival times. Numerous proposals have been made over the years, none of them unanimously accepted (for  exhaustive reviews of the problem, see \cite{Muga2000,Dassreview,DasNoth2021,Dasdas}).\\
\indent Remarkably, within the framework of de Broglie Bohm (dBB) theory \cite{Holland,BohmHiley} also called Bohmian mechanics - which is an alternative deterministic interpretation of quantum mechanics that re-establishes the notion of trajectory for particles - it is possible to define unambiguously the arrival time of a quantum particle at any point in space based on the precise calculation of the trajectory passing through that point \cite{Leavens1993,Leavens1996,Leavens1998,Finkelstein1999,Muga2000}.\\ 
\indent However, one of the problems associated with this dBB definition of arrival times concerns its link with the notion of quantum observable and POVM. Although the dBB definition is well-suited to the far-field regime, where it allows us to recover and justify standard results used in particle collision physics, it generally leads to difficulties in the near-field regime.  In particular, it has been shown that the dBB definition leads to fundamental problems when the flow of particles across a surface is associated with the phenomenon of `backflow' \cite{Muga2000}. 
In this  backflow regime (for reviews and general discusions see \cite{Allcock1969,Kijowski1974,Mielnik1994,Muga1999,Berry2010,Villanueva2020}), the same Bohmian trajectory  crosses a predefined detection spatial zone $D$  several times (i.e. at different times) from different sides \cite{Leavens1993}.   The arrival time is therefore not uniquely defined, and we must add a condition on the first passage, the second passage etc... of the particle in the detection region $D$ \cite{Leavens1995,Oriols1996}. Furthermore, the probability distribution of (first) arrival times given by dBB theory depends on the probability current $\mathbf{J}^\Psi(\mathbf{x},t)$ (more precisely on  the so called `truncated' probability current distribution \cite{Rafsanjani2023} associated with this multiplicity of passages on the detector). However, it has been shown \cite{Vona2013,Vonabis2013} that, in general, the probability current $\mathbf{J}^\Psi(\mathbf{x},t)$ is not associated with a POVM due to the presence of backflow. Since the notion of POVM is generally accepted as the most accurate theoretical description of a quantum observable, it seems apriori impossible to consider dBB arrival times as generally measurable. However, since these backflow phenomena are usually confined to interference zones or near-field regions of very limited spacio-temporal extension, it is generally accepted that observing this regime would be very difficult and has not yet been achieved.\\
\indent More fundamentally, the notion of a Bohmian arrival or travel time is still very controversial. For example it was claimed  that (in the context  of tunneling times) `\emph{Bohm's theory can make a definite prediction when standard quantum mechanics can make none at all}' \cite{Cushing1995} (similar speculations were discussed in \cite{Holland} p. 215 and \cite{Cushing1994} pp.53--55). This is a very strong statement which, if justified, would break the empirical equivalence between dBB mechanics and orthodox quantum mechanics. This would give a strong advantage to dBB theory and of course it generates controversy (see the discussion in \cite{Oriols1996}). One other controversy is perhaps that the probability current $\mathbf{J}^\Psi(\mathbf{x},t)$ is not unambiguously defined \cite{Finkelstein1999} (we can always add a term $\boldsymbol{\nabla}\times \mathbf{F}(\mathbf{x},t)$ to $\mathbf{J}^\Psi(\mathbf{x},t)$ without altering the local conservation law $\partial \rho^\Psi+\boldsymbol{\nabla}\cdot \mathbf{J}^\Psi=0$). There is thus a form of underdetermination concerning the uniqueness of dBB dynamics \cite{Oldofredi2024}. As a result, the physical meaning to be attributed to these arrival times based on the probability current is questionable. Nevertheless, there is a strong consensus concerning the far-field regime where the dBB trajectories are reduced to straight lines and which corresponds to the  scattering regime without backflow \cite{Daumer1996,Teufel} (in this regime a nice agreement between time of flight measurements and dBB  prediction has been recently analyzed for a double-slit experiment \cite{Das2022}).  It should be noted in this context, that some authors \cite{Ali,Moussavi,Home} oppose this indetermination of the Bohmian dynamics on the basis of the relativistic extension of the dBB theory using the Dirac equation, and Holland's work \cite{Holland} showing that Lorentz invariance fixes the arbitrariness in the form of the current.  \\  
\indent Recently, however, it has been proposed by Siddhant Das and Detlef D\"urr (hereinafter DD) to use the dBB approach for arrival times within the framework of the Pauli equation (i.e., the non-relativistic limit of the Dirac equation) for particles with spin$-1/2$ \cite{DasDurr2018,DasDurr2019}. Going far beyond previous works \cite{Ali,Moussavi,Home,Shadi,Chalinor,Melloy1998} based on the Dirac equation, the authors have defined a precise regime, in principle physically attainable, where the presence of backflow is not confined to the near-field domain. Importantly they found  a spin dependent distribution of arrival time $\mathcal{P}_{\textrm{dBB}}^{\Psi_{\hat{\mathbf{s}}}}(\Sigma,\tau)$ with interesting consequences for backflow.  This, of course, reopens the debate on the observation of Bohmian arrival times \cite{SA}.\\   
\indent Moreover, in a recent comment \cite{GTZ} Goldstein Tumulka and Zangh\`{i} (hereinafter GTZ) critically assessed the proposition and calculations made in  \cite{DasDurr2018,DasDurr2019} and showed that if DD results were exact, and if the dBB  first-arrival time spin dependent probability distribution could be identified with a POVM, then a contradiction occurs. Therefore, GTZ conclude, something must be wrong in the predictions given in \cite{DasDurr2018,DasDurr2019} concerning the observation of dBB arrival times.\\
\indent  Yet, as Das and Aristarhov stressed  in a reply \cite{DasAris},  DD never actually claimed that their proposition for a  dBB first arrival time measurement  was reduceable to a POVM, and thus far from contradicting \cite{DasDurr2018,DasDurr2019} the results of GTZ \cite{GTZ} only show that indeed the Bohmian first arrival time probability distribution  cannot be associated with a standard POVM. Therefore,  the real question  asked by GTZ and DD is whether or not `\emph{the statistics of the outcomes of any quantum experiment are governed by a POVM}' \cite{GTZ} and only by POVM.\\
\indent Here, to answer this question, we assess the analyses done by GTZ and DD. We show that while mathematically correct GTZ  conclusions are physically unjustifed. In particular, we emphasize that GTZ too strong reliance on POVM (which can be summarized by the slogan `POVM and only POVM') is mostly a prejudice of the orthodox theory of quantum measurements that must be generally abandonned in the light of the dBB theory.  As we show, although POVMs are necessary, they are not sufficient to describe a Bohmian measurement process. Moreover, we also stress that contrarily to DD claims experimental observation of the first arrival time probability distribution requires explicit consideration of the detector physics during the measurement process. Three key messages emerge from our analysis: First (i), in agreement with POVM no-go theorems, there is no universal detector for arrival time. Second (ii), every arrival time detector built for working at time $t$ is in general a very invasive device and could prohibit subsequent detections at later time $t'>t$ (even if the  measurement at time $t$ didn't actually occur because the detector didn't fire; this is an instance of negative-result quantum measurements). In other words, in the dBB framework one must distinguish between probability of being here at time $t$ and probability of being detected here at time $t$. Finally (iii),  first arrival times are defined within the dBB framework and as such require a post selection of the data: The whole procedure is thus theory laden. In the end we show that when all these features are taken into account nothing prohibit the experimental  observation of the first arrival time probability distribution predicted by DD.\\
\indent Moreover, as a follow up of the previous studies by DD it must be mentionned that the philosopher Tim Maudlin has in various occasions on social media \cite{Maudlin1,Maudlin2,Maudlin3} discussed the possibility to use the results obtained by DD  for the spin  dependent time arrival probability distribution  in order to develop new dBB-based faster-than-light communications protocols involving pairs of entangled  spin$-1/2$ particles. This `Bell telephone' possibility clearly contradicts the no-signalling theorem deduced from quantum mechanics in the context of Bell's theorem. More precisely, we demonstrate in quantum field theory that local commutativity and microcausality impose this no-signalling constraint.\cite{nonsig,nonsigBell}  As stressed by Bell: `\emph{It is as if there is some kind of conspiracy, that something is going on behind the scenes which is not allowed to appear on the scenes}'. \cite{Bell2} In fact, dBB theory emphasizes the crucial role of Born's rule in this derivation as shown by Valentini \cite{nonsigValentini}, and Born's rule is fundamentally linked to the existence and use of POVMs in quantum mechanics. Unless we relax Born's rule, i.e. abandon `quantum equilibrium',\cite{nonsigValentini} or modify quantum mechanics, it is thus impossible in the quantum framework to exploit the violation of Bell's inequalities (i.e., the nonlocality of dBB theory) to transmit a signal faster than light. Therefore, in a recent extension of their original comment, GTZ stress  \cite{GTZ2} that DD results, assuming Born's rule, also strongly contradict the no-signalling theorem and therefore  conflict with standard quantum mechanics.\\
\indent  However, we demonstrate in this article that a new analysis of the problem, in particular in relation to point iii) above, can remove the paradoxes. In fact, according to our analysis, it becomes possible to measure the probability distribution predicted by DD without violating the no-signalling theorem, thereby ruling out the possibility of supraluminal transmission channels contradicting Bell's theorem.\\  
\indent The layout of our article is the following: In the next section we briefly review   dBB theory and show,  with the help of one typical example, how  it allows us to give physical answers to questions that look devious in the orthodox quantum theory. In Section \ref{section2} we analyse the nature of measurements in dBB theory and stress the limitation of the notion of POVM. In Section \ref{section3} we review the arrival time problem in the dBB theory and discuss DD and  Maudlin proposals aswell as the counter analysis  by GTZ. In Sections \ref{section4}, \ref{section5} and \ref{section6} we discuss the theory of detectors and the impact this has on dBB theory. In particular we define regimes of strong and weak coupling for detections.    Finally, in the last Section \ref{section7} we resume the first arrival time problem in the dBB theory  and DD's proposal involving the Pauli or Dirac equation for particle with spin$- 1/2$ and show how we can in principle measure the distribution predicted  by DD.  
\section{Bohmian inference}
\label{section2}
The concept of experimental measurement, and more precisely of so-called direct experimental measurement, has always been a source of debate in physics since its foundation. Take for example Rutherford's experiment where a beam of $\alpha$ particles passes through a thin film of gold. From the deviation of $\alpha$ particles at high angles (i.e., in backscattering) Rutherford deduced (or rather induced) the existence of atomic nuclei acting as very compact centers of diffusion. Seen in hindsight, however, it is impossible to deduce a theory from this Rutherford experiment. As Albert Einstein understood perfectly well, the best we can say or infer  is that having a theory we can define what is measurable, or not, and then compare the predictions to the results. In other words, any measurement is necessarily indirect and presupposes a theoretical model. As he explained in 1926 to Heisenberg who claimed to be able to build a quantum theory by limiting himself only to what is observable:
\begin{quote}
`\emph{it is the theory which alone decides what is measurable}'.\cite{Heisenberg}
\end{quote} This is the heart of the hypothetico-deductive method!

In quantum physics it is the forgetting of this elementary truth concerning the scientific method which is responsible for numerous errors of interpretation. So, let's take Young's famous two-slit experiment. According to Bohrian quantum doxa it is impossible to interpret the observation of interference fringes using the concept of a continuous trajectory followed by individual particles. For Bohr and Heisenberg for example this would indeed amount to saying that the trajectory of a particle passing through hole A is influenced by the existence of hole B through which it however did not pass! From a classical perspective this is a priori nonsense. But all this shows is that certain classical `Newtonian' prejudices oppose a simple interpretation of Young's slit experiment in terms of trajectory.

However, we know that the pilot-wave dBB theory developed by Louis de Broglie in 1927 \cite{deBroglie1927,Valentini2009,deBroglie1930} and rediscovered by Bohm in  1952 \cite{Bohm1952a,Bohm1952b}  makes it possible to precisely explain this interference experiment using trajectories \cite{BohmHiley,Holland}. In this dBB theory the trajectories of the particles are strongly curved by the presence of potentials of a specifically quantum nature which free themselves from the overly Newtonian prejudices of Bohr and Heisenberg. We can notice that the Newtonian reading made by Heisenberg and Bohr is biased. Indeed, in his seminal work \textit{Opticks} published in 1704 \cite{Newton}, Newton sought to explain the observation of diffraction and Newton's rings using a “access theory” involving forces acting on the particles of light, and which, in many aspects, anticipates the notion of Bohmian quantum potential. In turn, this quantum theory of dBB, giving meaning to the notion of trajectory, makes it possible to define and characterize what is a `good' measurement, or not, in complete agreement with Einstein's hypothetico-deductivism.

To be more precise we remind that the dBB velocity of a particle is given in the simplest non-relativistic spinless theory  by the de Broglie guidance formula: \cite{BohmHiley,Holland}
\begin{eqnarray}
\frac{d}{dt}\mathbf{x}(t):=\mathbf{v}^\Psi(\mathbf{x}(t),t)=\frac{\mathbf{J}^\Psi(\mathbf{x}(t),t)}{|\Psi(\mathbf{x}(t),t)|^2}=\textrm{Im}[\frac{\boldsymbol{\nabla}\Psi(\mathbf{x}(t),t)}{m\Psi(\mathbf{x}(t),t)}]=\frac{\boldsymbol{\nabla}S^\Psi(\mathbf{x}(t),t)}{m}\label{1}
\end{eqnarray}   where $\mathbf{J}^\Psi$ is the probability current,  $m$ the particle mass, and $S^\Psi$ the phase of $\Psi$ (i.e., the quantum Hamilton-Jacobi action).      
This first-order differential equation $\frac{dx}{v^\Psi_x}=\frac{dy}{v^\Psi_y}=\frac{dz}{v^\Psi_z}=dt$ can be integrated (at least numerically) and defines the Bohmian trajectories of the particle. In particular, trajectories obtained from this dynamics can in general not cross.\\
\indent A important feature of the dBB theory concerns probability and statistics. Indeed, from the law of conservation and the definition of the probability current the dBB theory shows that if an ensemble of similarly prepared particles are statistically $\rho^\Psi_{t_0}:=|\Psi|^2(\mathbf{x}(t_0),t_0)$ distributed at an initial time $t_0$ this will be so at any other time $t$: $\rho^\Psi_{t}:=|\Psi|^2(\mathbf{x}(t),t)$. In other words, from this property called equivariance, Born's rule  $\rho^\Psi:=|\Psi|^2$ is naturally consistent with the dBB theory and therefore the statistical predictions of standard quantum mechanics can be recovered within this framework.\cite{BohmHiley,Holland}\\   
\indent Moreover, in the double-slit experiment all this has huge consequences. Consider the case of a single electron wave function 
\begin{eqnarray}
\Psi(x,y,z,t)=\Psi_0(x-a/2,y,z,t)+\Psi_0(x+a/2,y,z,t)\nonumber\\ \label{2}
\end{eqnarray} where $\Psi_0(x,y,z,t)$ is a propagating wavepacket initially  (i.e. at time $t=0$) centered on the origin and subsequently propagating along the $z$ direction while it also spreads. Assuming $\Psi_0(x,y,z,t)=\Psi_0(-x,y,z,t)$ we thus deduce from  Eqs.~\ref{1} and \ref{2} that dBB trajectories cannot cross the symmetry plane $x=0$. Furthermore, suppose that at time $t=0$ the wave function  $\Psi_0(x,y,z,t=0)$ has a finite spatial support $\Delta_0(\mathbf{0})$ such that the two wave packets  $\Psi_0(x-a/2,y,z,t=0)$ and $\Psi_0(x+a/2,y,z,t=0)$ are not overlapping (i.e., $\Delta_0(a/2,0,0)\cap\Delta_0(-a/2,0,0)=\emptyset$). Thus, dBB theory allows us to retrodict: if we record the particle at the plane $z$ in the zone $x>0$ we can indeed retrodict that the particle was necessarily coming at $t=0$ from the wavepacket located in the upper side of the screen $z=0$ i.e. centered on $x=+a/2$. The converse is true for a particle detected in the region $x<0$ allowing us to infer that it was coming from the lower wavepacket centered on $x=-a/2$.

With the dBB theory the probability for the particle detected at time $t$ to come from the wave packet centered on $x=a$ is thus rigorously 
\begin{eqnarray}
\mathcal{P}^{\Psi}_{+a}(t)=\int_{x\geq 0} d^3\mathbf{x}|\Psi(\mathbf{x},t)|^2=\int_{\mathbf{x}-\mathbf{\hat{x}}\frac{a}{2}\in\Delta_0} d^3\mathbf{x}|\Psi_0(x-a/2,y,z,t=0)|^2=\frac{1}{2}
\end{eqnarray}   with a similar an symmetric expression for $\mathcal{P}^{\Psi}_{-a}(t)$. 
This is easily deduced from the two  properties i) the dBB trajectories cannot cross and, ii) the conservation of the probability fluid is preserved along trajectories, i.e., $\rho^{\Psi}(\mathbf{x}(t),t)\delta^3\mathbf{x}(t)=\rho^{\Psi}(\mathbf{x}(t=0),t=0)\delta^3\mathbf{x}(t=0)$. According to dBB theory we also have $\rho^{\Psi}(\mathbf{x}(t),t)=|\Psi|^2(\mathbf{x}(t),t)$ a probabilistic rule that was assumed by de Broglie even before Max Born!\\ 
\indent For the present discussion we emphasize that we can write 
\begin{eqnarray}
\mathcal{P}^{\Psi}_{\pm a}(t)=\langle \Psi(t)|\hat{O}_{\pm a}|\Psi(t)\rangle \label{4}
\end{eqnarray} 
with the operators $\hat{O}_{+a}=\int_{x\geq 0} d^3\mathbf{x}|\mathbf{x}\rangle\langle \mathbf{x}|$, $\hat{O}_{-a}=\int_{x\leq 0} d^3\mathbf{x}|\mathbf{x}\rangle\langle \mathbf{x}|$  defined as sums of projectors, i.e., a special case of POVM (in dBB mechanics every measurements 	are ultimately analyzed in terms of spatial projections).\\
\indent We briefly remind that mathematically speaking POVMs $\hat{O}_n$ are linear self-adjoint operators (i.e., $\hat{O}_n=\hat{O}^\dagger_n$) acting on a Hilbert space $\mathcal{H}$, such that $\sum_n\hat{O}_n=\hat{I}$. These operators obey the positivity condition $\hat{O}_n>0$ which actually reads $\langle\Psi|\hat{O}_n|\Psi\rangle>0$ whatever the state $|\Psi\rangle\in \mathcal{H}$. This last condition is naturally needed  in order to interpret $\langle\Psi|\hat{O}_n|\Psi\rangle$ as a probability. We note that, rigorously speaking, a POVM denotes the set of all linear operators satisfying the previous conditions.  By extension, it's common to call any member of the preceding family a POVM, and we'll continue to use this convention hereafter. The theory of quantum measurement ultimately relies on the concept of POVM (for an introduction to POVM and its use in quantum information processing see \cite{Peres,Brandt} and for a more general and precise discussion related to the measurement process and  dBB theory see \cite{Durr,Vonabis2013,DGZ,Tumulka}). These operators indeed constitute fundamental mathematical tools formalizing the generalized von Neumann  measurements coupling a system $S$ to a pointer $M$.   Moreover, in the DBB theory relying on spatial measurements   the fundamental role is played by projectors $|q\rangle\langle q|$ (where $q$ is sa coordinate vector in the configuration space of the system). We include in the Appendix \ref{appendix0} a brief description of POVM measurement formalism applied to dBB theory for particles. This approach is non ambiguous at least in the non relativistic regime for particle with or without spin  an by extension for Dirac (fermionic) relativistic particles (the extension to bosonic quantum fields is also possible but relies on different beables or hidden variables  than particle positions $q$ and will not be considered here). \\
\indent  Going back to our previous example  with the double-slit experiment it is central to observe that while $\mathcal{P}^{\Psi}_{\pm a}(t)$ are obtained from standard POVMs the Bohmian algorithm to interpret these experimentally observable quantities as physical properties associated with the system at time $t=0$ doesn't work for an arbitrary wave function.  Indeed, the previous example strongly relies of the symmetry of $\Psi_S$. For a different superposition  (for example  by adding a phase:  $\Psi'_S(x,y,z,t)=\Psi_0(x-a/2,y,z,t)+e^{i\chi}\Psi_0(x+a/2,y,z,t)$) the interpretation of $\langle \Psi'(t)|\hat{O}_{+a}|\Psi'(t)\rangle$  as a probability $\mathcal{P}^{\Psi'}_{+a}$ for the particle to be initially in the upper wave packet will not generally hold! It will however work for the cases $\chi=0$ or $\pi$.\\ 
\indent In other words, in general Eq.~\ref{4} doesn't define  genuine Bohmian `which-path' observables. Moreover for an arbitrary wave function $\Psi(x,y,z,t)=\alpha\Psi_0(x-a/2,y,z,t)+\beta\Psi_0(x+a/2,y,z,t)$ it will be possible to define other POVMs 
\begin{eqnarray}
\hat{O}^{(\Psi)}_{\pm a}=\int_{\mathbf{x} \in \Delta^{(\Psi)}_{\pm a}} d^3\mathbf{x}|\mathbf{x}\rangle\langle \mathbf{x}|
\end{eqnarray}
where $\Delta^{(\Psi)}_{\pm a}$ are spatial domains image of $\Delta_0(\pm a,0,0)$ through the Bohmian flow $\mathbf{x}(t)=\mathbf{F}^{(\Psi)}_t(\mathbf{x}_0,t=0)$, i.e., $\Delta^{(\Psi)}_{\pm a}=F^{(\Psi)}_t(\Delta_0(\pm a,0,0),t=0)$. In the dBB formalism this can written
\begin{eqnarray}
\mathcal{P}^{\Psi}_{\pm a}(t)=\int\mathbb{I}_{\Delta^{(\Psi)}_{\pm a}}(\mathbf{x}(t))\rho^{\Psi}(\mathbf{x}(t),t)d^3\mathbf{x}(t)\\ \nonumber
=\int\mathbb{I}_{\Delta^{(\Psi)}_{\pm a}}(\mathbf{x}(t))\rho^{\Psi}(\mathbf{x}_0,t=0)d^3\mathbf{x}_0
\end{eqnarray}
where $\mathbb{I}_{\Delta^{(\Psi)}_{\pm a}}(\mathbf{x}(t))=\int_{\mathbf{u} \in \Delta^{(\Psi)}_{\pm a}} d^3\mathbf{u}\delta^3(\mathbf{x}(t)-\mathbf{u})$ is an indicator function  
\\
\indent In fact, all this can be interpreted in another way: An experimenter content with measuring the spatial distribution of particle arrival on the detection screen will generally not be able to trace the notion of path followed and thus obtain `which-path' type information without performing a post-analysis on the events detected. So, in our example, the experimenter will be able to post-select the events detected in the x-positive region in order to obtain physical information. It's the theory, in this case that of de Broglie Bohm, that makes it possible to interpret and give meaning to the raw data. \\
\indent It is clearly an example of Einstein's credo `the theory decides what is to be measured'!\\
\indent Having explained this, we are now ready to discuss the relationship between the concept of arrival time in Bohmian mechanics and the notion of POVM.
\section{Can we observe Bohmian first arrival time ? (first round)}
\label{section3}
\indent The notion of arrival-time can be precisely defined in dBB theory. Consider a region of space, say a $\Sigma$ surface, then for a given wave function $\Psi_t$ the dBB trajectories $\mathbf{x}(t)$ arbitrarily integrated from an initial time $t_0=0$ and passing through this surface define the successive arrival times of the particle on this surface. In general, these times are not unique, as the particle can zig-zag around $\Sigma$.  In cases where we can define a dBB first instant of arrival $\tau^\Psi_\Sigma$ on $\Sigma$ (which is generally true for time-dependent problems where the wave function $\Psi_t$is non-stationary), we formally write: \cite{Daumer1996,Teufel,DasDurr2018,DasDurr2019,Das2022,Rafsanjani2023}
\begin{eqnarray}
\tau^\Psi_\Sigma=\textrm{inf}\{t:\mathbf{x}(t)\in \Sigma\}
\end{eqnarray}
\indent The distribution of arrival-time is generally obtained from the  probability current $\mathbf{J}^\Psi(\mathbf{x},t)$ projected onto the detection surface element having the direction $\mathbf{n}(\mathbf{x})$. Considering an infinitessimal surface $d\Sigma_\mathbf{x}$ the number or particles crossing this surface during an infinitessimal interval of  time $\delta t$ around $t$ is given by
 \begin{eqnarray}
\mathcal{P}_{\textrm{dBB}}^\Psi(\mathbf{x},t)\delta t :=|\mathbf{J}^\Psi(\mathbf{x},t)\cdot\mathbf{n}(\mathbf{x})|d\Sigma_\mathbf{x}\delta t 
=\rho^{\Psi}(\mathbf{x}(t),t)\delta^3\mathbf{x}(t)=\rho^{\Psi}(\mathbf{x}(t_0),t_0)\delta^3\mathbf{x}(t_0)\label{dBBflux}
\end{eqnarray} where we used the conservation of probability in the second and third lines with the volume $\delta^3\mathbf{x}(t)=|\mathbf{v}^\Psi(\mathbf{x},t)\cdot\mathbf{n}(\mathbf{x})|d\Sigma_\mathbf{x}\delta t$ and where $\mathbf{x}(t_0)$ are the initial coordinates for   the dBB trajectory connected to $\mathbf{x}(t)$. 
Moreover, if we consider only what is occuring in this time window $\delta t$ then $\mathcal{P}_{\textrm{dBB}}^\Psi(\mathbf{x},t)=\rho^{\Psi}(\mathbf{x}(t_0),t_0)\delta^3\mathbf{x}(t_0)\frac{1}{\delta t}$ can be interpreted as an arrival time distribution for the elementary surface $d\Sigma$. Integrating over the surface $\Sigma$ and postselecting only on those events coresponding to a first arrival  the probability distribution of first arrival reads 
\begin{eqnarray}
\mathcal{P}_{\textrm{dBB}}^\Psi(\Sigma,\tau)=\int_\Sigma|\mathbf{J}^\Psi(\mathbf{x},\tau=\tau^\Psi_\Sigma)\cdot\mathbf{n}(\mathbf{x})|d\Sigma_\mathbf{x}=\int_V \delta(\tau-\tau^\Psi_\Sigma)\rho^{\Psi}(\mathbf{x}_0,t_0)d^3\mathbf{x}_0
\end{eqnarray} where $\mathbf{x}_0=\mathbf{x}(t_0)$. These two very general expressions are equivalent and were used by DD \cite{DasDurr2018,DasDurr2019,Das2022} partly based on previous works by Leavens \cite{Leavens1993,Muga2000,Hannstein2005} and D\"{u}rr et collaborators \cite{Daumer1996,Teufel} (for other related works see  \cite{Rafsanjani2023,Gruebl2023,Beau2024}). 

We stress that the absolute value is required since the particle can come from the wrong side in presence of back flow.
In this regime, the probability current is backpropagating, i.e., $\mathbf{J}^\Psi(\mathbf{x},t)\cdot\mathbf{n}(\mathbf{x})<0$ even if the incident wavepacket  $\Psi$ contains only propagative components $\Psi_k$ (i.e. plane waves) which separately satisfy $\mathbf{J}^{\Psi_k}(\mathbf{x},t)\cdot\mathbf{n}(\mathbf{x})>0$ . In other words, backflow can be seen as an interference effect specific of wave mechanics. We also emphasize that in general $\mathcal{P}_{\textrm{dBB}}^\Psi(\Sigma,\tau)$ is not normalized, i.e., $\int d\tau\mathcal{P}_{\textrm{dBB}}^\Psi(\Sigma,\tau)\leq 1$, because not every trajectories are necessarily crossing $\Sigma$.\\
\indent  In usual scattering experiments where the incident wave packet $\Psi_{t_0}$ is well localized in space and where the detection  surface  is located in the far-field (the far-field is the regime where  $r\gg \lambda$, with   $r$ a typical distance between the source and the detector   and $\lambda$ a typical wavelength) we can completely neglect back flow. In this regime, the distribution of first arrival times becomes the distribution of arrival times altogether. \cite{Daumer1996,Teufel} There's no longer any need to involve post-selection on arrival times, and the probability distribution reduces to the standard formula used in collision physics (e.g., for evaluating scattering cross-sections), regardless of any knowledge of Bohmian theory. We emphasize that semiclassical and far-field regimes are often used  in the orthodox quantum interpretation but these approximations appear only as limiting special cases of the dBB framework situations where trajectories are classical-like (i.e., because the quantum potential is negligible). In the dBB framework all kind of vagueness concerning classicality can be easily removed and the physical interpretation of $\mathcal{P}_{\textrm{dBB}}^\Psi(\Sigma,\tau)$ is non ambiguous even in regimes where the far-field positivity condition $\mathbf{J}^\Psi(\mathbf{x},t)\cdot\mathbf{n}(\mathbf{x})>0$ doesn't hold anymore. The ontological clarity of classical physics is thus recovered even in the quantum regime! \\
\indent Yet, the fact that the probability $\mathcal{P}_{\textrm{dBB}}^\Psi(\Sigma,\tau)$  can be mathematically constructed from the notion of Bohmian trajectories does not explain how this probability can be measured. Indeed quantum mechanics is highly contextual and one should clearly distinguish the probability of being from the probability of being detected.   In fact, it is well accepted that in the far-field regime, i.e., in the absence of back flow, the $\mathcal{P}_{\textrm{dBB}}^\Psi(\Sigma,\tau)$  distribution is directly measurable, in line with results obtained in studies of scattering and collision processes. This is also what emerges from the observation of diffraction and interference phenomena, also observed in the far-field and in very good agreement with the predictions for $\mathcal{P}^\Psi(\Sigma,\tau)$  given by Bohmian theory.\\
\indent   Nevertheless, nothing is less certain for the more general regime where the back-flow phenomenon is predicted by theory. The measurability of $\mathcal{P}_{\textrm{dBB}}^\Psi(\Sigma,\tau)$ can then be questioned. This is exactly the regime considered by DD \cite{DasDurr2018,DasDurr2019} in a situation involving a spin-1/2 particle exhibiting a back-flow-like phenomenon even quite far from the source. To justify the measurability of  $\mathcal{P}_{\textrm{dBB}}^\Psi(\Sigma,\tau)$  in this regime, DD points out that a precise theory of detection is by no means necessary to understand the far-field regime in very good agreement with the theory. Similar statements were given by  D\"{u}rr and Teufel  in a known textbook:
\begin{quote}
`\emph{We should follow the common practice of quantum physics and henceforth not worry about the presence of detectors, simply taking for granted that the detection is designed in such a way that it does not mess up the trajectories too much}' \cite{Teufel} p. 347.
\end{quote}In DD view, the same could be expected in the new situation, even in the presence of back-flow. They wrote: 
\begin{quote}
`\emph{We expect that in the experiment proposed in this paper the same will be true, i.e., the detection event should not be drastically disturbed by the presence of the detector. }'.\cite{DasDurr2018}
\end{quote}However, it is not difficult to show that this necessarily leads to difficulties and even paradoxes such as those highlighted by GTZ \cite{GTZ,GTZ}.\\
\indent To keep the description of the situation described by DD as simple as possible, we recall that it considers a spin-1/2 particle confined in a cylindrical guide with symmetry axis $z$. Initially, the particle is described by a  strongly localized wave function 
\begin{eqnarray}
\Psi_{\hat{\mathbf{s}}}(\rho,z,t_0)=\chi_{\hat{\mathbf{s}}}\cdot\Phi(\rho,z,t_0)
\end{eqnarray} where $\chi_{\hat{\mathbf{s}}}$ is a two component spinor such  that $\chi_{\hat{\mathbf{s}}}^\dagger \boldsymbol{\sigma}\chi_{\hat{\mathbf{s}}}=\hat{\mathbf{s}}$ ($\hat{\mathbf{s}}$ is a unit vector defining the spin direction and  $\boldsymbol{\sigma}=[\sigma_x,\sigma_y,\sigma_z]$ are the Pauli matrices). Initial particle spatial confinement along the $z$ direction is provided by a potential well. When this is removed on one side only, the wave packet moves towards $z>0$ while preserving the structure of the spinor $\chi$, which remains unchanged. The wave function then becomes $\Psi_{\hat{\mathbf{s}}}(\rho,z,t)=\chi_{\hat{\mathbf{s}}}\cdot\Phi(\rho,z,t)$ the spatial part, preserving its rotation invariance over time. The dBB theory applied to the Pauli equation leads to a probability current along the $z$ direction: 
\begin{eqnarray}
J_z^{\Psi_{\hat{\mathbf{s}}}}(\mathbf{x},t)=|\Phi(\rho,z,t)|^2\frac{\partial_zS(\rho,z,t)}{m}+\frac{\hat{\mathbf{s}}\cdot\hat{\boldsymbol{\varphi}}}{2m}\partial_\rho|\Phi(\rho,z,t)|^2.
\label{currentspin}\end{eqnarray} In this formula   the first term is a convective current reminiscent from the formula used for a spinless particle ($S$ is the phase of the wave packet $\Phi(\rho,z,t)$).  The second term is a spin current associated with the magnetic structure of the electron ($\hat{\boldsymbol{\varphi}}$ is a unit vector for the polar angle direction). We stress that Eq.~\ref{currentspin} is an application of the  non-relativistic Gordon formula $\mathbf{J}^\Psi=\frac{1}{2mi}(\Psi^\dagger\stackrel{\textstyle\leftrightarrow}{\rm \boldsymbol{\nabla}}\Psi)+\frac{1}{2m}\boldsymbol{\nabla}\times[\Psi^\dagger \boldsymbol{\sigma} \Psi]$ for the probability current of an electron. In the relativistic regime it is more convenient to use Dirac current $\mathbf{J}^\Psi=\Psi^\dagger\boldsymbol{\alpha}\Psi$  using the bispinor $\Psi$ (for previous studies using the Dirac equation see \cite{Ali,Moussavi,Home,Shadi,Chalinor}).\\
\indent Two regimes are clearly identifiable. Firstly, in the longitudinal case where the spin vector is aligned with the $\pm z$ direction, only the convective current survives.  The dBB theory then gives the same trajectories as for a spinless particle, and in particular the absence of back-flow.  The second regime is more interesting and corresponds to the case of a purely transverse spin in the $\pm x$ direction, for example. In this case, the spin current reads $\pm\frac{\cos{\varphi}}{2}\partial_\rho|\Phi(\rho,z,t)|^2$ and can clearly change sign. In the configuration considered by DD, the spin current can more than compensate for the positive convective current, and so in some cases we get a back-flow  $J_z^{\Psi_{\hat{\mathbf{s}}}}<0$. \\
\indent  Using these predictions for the probability current, we can construct probability distributions for the first arrival times on a given cross section $\Sigma$ of the wave guide at $z=const>0$ in both longitudinal and transverse spin configurations. The distribution $\mathcal{P}_{\textrm{dBB}}^{\Psi_{\pm\hat{\mathbf{z}}}}(\Sigma,\tau)$ for the longitudinal case is similar for both $\pm z$ possibilities (i.e., $\mathcal{P}_{\textrm{dBB}}^{\Psi_{\hat{\mathbf{z}}}}(\Sigma,\tau)=\mathcal{P}^{\Psi_{-\hat{\mathbf{z}}}}(\Sigma,\tau)$). Qualitatively, the distribution starts from zero for $\tau=0$, approaches a maximum, then slowly decreases to zero for $\tau$ tending towards infinity. This probability distribution gives the same result as for the spinless particle case. The transverse configuration is more surprising. We have first a rotational invariance  $\mathcal{P}_{\textrm{dBB}}^{\Psi_{\hat{\mathbf{s}}}}(\Sigma,\tau)=\mathcal{P}_{\textrm{dBB}}^{\Psi_{\hat{\mathbf{s}}'}}(\Sigma,\tau)$  for any choice of the transverse spin vector (for example $\hat{\mathbf{s}}=+\hat{\mathbf{x}}$  or $\hat{\mathbf{s}}'=-\hat{\mathbf{x}}$) which was expected based on the cylindrical symmety of the problem. Qualitatively, the probability distribution for the transverse case resembles the longitudinal one. The probability starts from zero at $\tau=0$, approaches a maximum and decreases. Here, however, the distribution is more peaked and the decay more pronounced. Remarkably, after a characteristic time $\tau_{max}$, the distribution rigorously cancels out and remains so for any time $\tau>\tau_{max}$.\\
\indent It is at this point that GTZ deduce a contradiction. Assuming that the distribution of arrival times is given by a POVM and that we have 
$\mathcal{P}_{\textrm{dBB}}^{\Psi_{\hat{\mathbf{s}}}}(\Sigma,\tau)=\langle \Psi_{\hat{\mathbf{s}}}|\hat{O}(\Sigma,\tau)|\Psi_{\hat{\mathbf{s}}}\rangle$ GTZ show it would imply  
\begin{eqnarray}
\langle \Psi_{\hat{\mathbf{z}}}|\hat{O}(\Sigma,\tau)|\Psi_{\hat{\mathbf{z}}}\rangle+\langle \Psi_{-\hat{\mathbf{z}}}|\hat{O}(\Sigma,\tau)|\Psi_{-\hat{\mathbf{z}}}\rangle
=\langle \Psi_{\hat{\mathbf{x}}}|\hat{O}(\Sigma,\tau)|\Psi_{\hat{\mathbf{x}}}\rangle+\langle \Psi_{-\hat{\mathbf{x}}}|\hat{O}(\Sigma,\tau)|\Psi_{-\hat{\mathbf{x}}}\rangle. \label{sum}
\end{eqnarray} However, from the above mentionned symmetries of the arrival time distribution that would imply 
$\mathcal{P}_{\textrm{dBB}}^{\Psi_{\hat{\mathbf{z}}}}(\Sigma,\tau)=\mathcal{P}^{\Psi_{\hat{\mathbf{x}}}}(\Sigma,\tau)$ which is in general  not true (in particular for $\tau_{max}$). Therefore, as shown by GTZ, the dBB first arrival time distribution cannot be identified with a POVM.\\
\indent This result is unavoidable and since any experimental quantum statistics are assumed to be represented by POVM this seems to imply that $\mathcal{P}_{\textrm{dBB}}^{\Psi_{\hat{\mathbf{}}}}(\Sigma,\tau)$ is not measurable.\\

\begin{figure}[h]
\includegraphics[width=9cm]{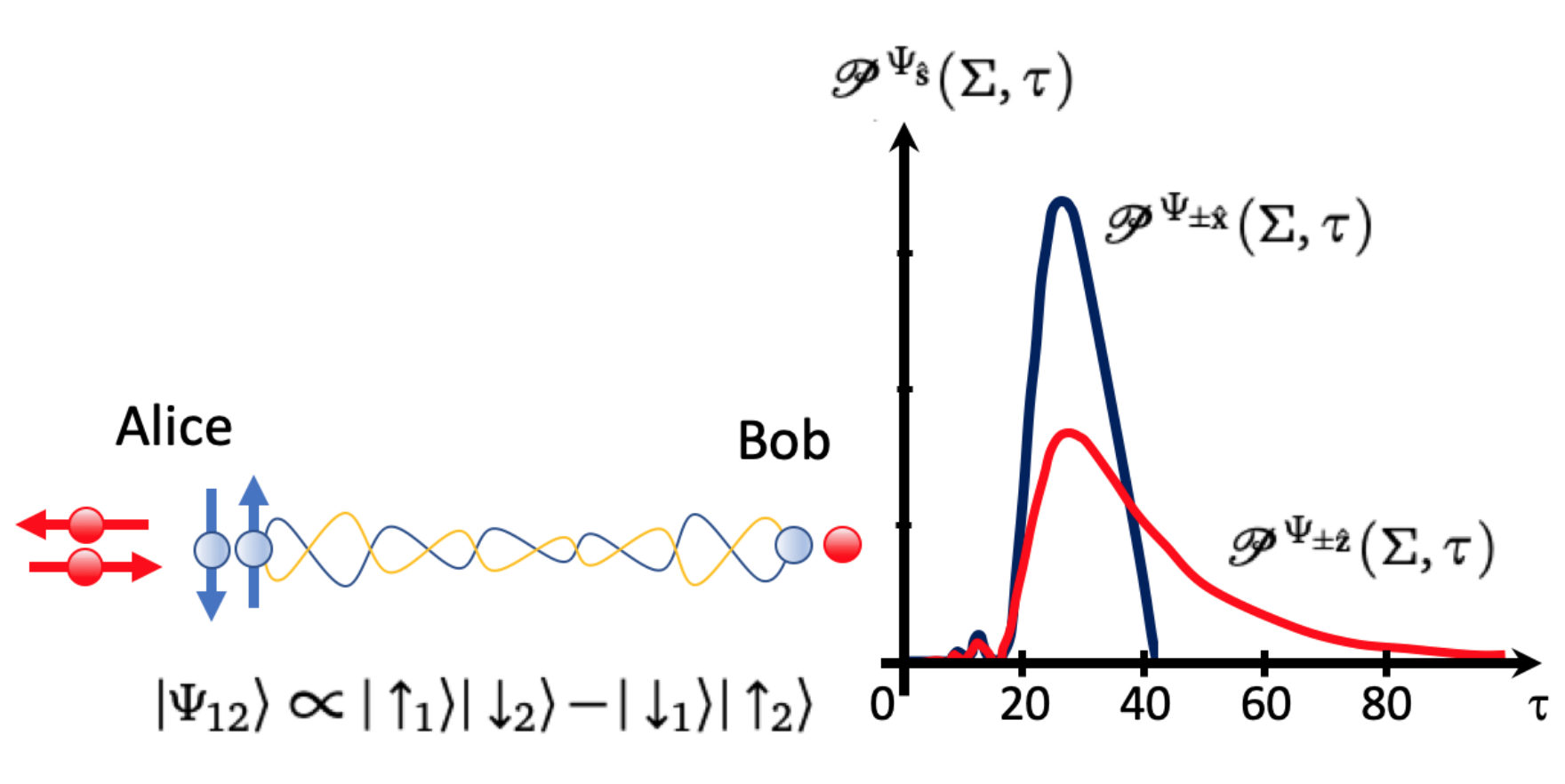} 
\caption{Principle of the experiment proposed by Das and Maudlin to build a faster than light Bell telephone. A pair of spin-$1/2$ particle in a perfectly entangled  EPR state is separated and analyszed by two agents Alice and Bob. Alice can measure the spin of her particle along the $\pm\hat{\mathbf{s}}$ unit directions. Specifically she considers the case $\pm\hat{\mathbf{z}}$ (longitudinal) and   $\pm\hat{\mathbf{x}}$ (transverse). On his side Bob measure the first arival time distribution $\mathcal{P}_{\textrm{dBB}}^{\Psi_{\hat{\mathbf{s}}}}(\Sigma,\tau)$ for his particle (still ignoring the spin of his particle). Let say he is measuring at a time $\tau\gg \tau_{max}$ at which the distribution $\mathcal{P}_{\textrm{dBB}}^{\Psi_{\hat{\mathbf{x}}}}(\Sigma,\tau)$ vanishes but $\mathcal{P}_{\textrm{dBB}}^{\Psi_{\hat{\mathbf{z}}}}(\Sigma,\tau)$ does not (the distributions are here taken and freely adapted from \cite{DasDurr2018}).  If Bob, in his remote lab, detects an event at $\tau\gg \tau_{max}$ then he can deduce that Alice was measuring her spin along the  longitudinal direction $\pm\hat{\mathbf{z}}$? Her measurement affects nonlocally the dBB dynamics of the particle detected by Bob: This is a form of faster than light communication contradicting Bell's no-signalling. \cite{nonsig}} \label{Fig0}
\end{figure}

\indent There's another very good reason justifying  this result. If $\mathcal{P}_{\textrm{dBB}}^{\Psi_{\hat{\mathbf{s}}}}(\Sigma,\tau)$ was a POVM, then we would have a way of beating the no-signalling theorem.  We could in fact use a Bell-type scenario with an entangled  pair of EPR spin-1/2 particles to send signals faster than light! The idea has been recently proposed by Maudlin  in several interviews referring to DD work \cite{Maudlin1,Maudlin2,Maudlin3}.  The point is that if we have an EPR-Bohm pair $|\Psi_{12}\rangle\propto|\uparrow_1\rangle|\downarrow_2\rangle-|\downarrow_1\rangle|\uparrow_2\rangle$ and if we send one of the particle to Pluto where Alice can measure the spin value along one arbitrary direction then Bob on earth could by recording the first arrival time distribution of the second particle (send into the wave guide proposed by DD) know instantaneously the value of the spin measured by Alice just by observing an event at time $\tau>\tau_{max}$ (see Figure \ref{Fig0}). Critically, this Maudlin-DD proposal is based on the assumption that the distribution of first arrival times is identifiable with a POVM i.e.,  $\langle \Psi_{12}|\hat{O}_2(\Sigma,\tau)|\Psi_{12}\rangle$. Moreover, POVM are central for deriving the no-signalling theorem in relativistic quantum mechanics and  this result is central to guaranteeing a peacefull relationship between quantum mechanics and Einstein's relativity theory. It is therefore apriori highly desirable that the $\mathcal{P}_{\textrm{dBB}}^\Psi(\Sigma,\tau)$ distribution is not a POVM, otherwise it would jeopardize all quantum field theory! Clearly Eq.~\ref{sum} agrees with the no-signalling theorem since the sum is independent of the spin basis chosen and Bob not knowing the result of Alice must  observe a random mixture.   Taken all together, this doesn't leave much hope for the measurability of the Bohmian distribution  $\mathcal{P}_{\textrm{dBB}}^\Psi(\Sigma,\tau)$. 

\indent Of course, there remains the possibility that DD's and Maudlin's results and analyses are correct, in which case it could be that performing the experiment could indeed defeat the POVM-based no-signalling theorem. This seems highly speculative, however, since this would imply a whole new physics beyond the standard theory of measurement based on POVMs and this would call into question the peaceful consensus between quantum mechanics and relativity theory. Therefore, it's likely that another, more nuanced answer is the right one.  In what follows, we shall show both that GTZ's criticisms are too strong and that DD (and Maudin) are too confident about the measurability of the $\mathcal{P}_{\textrm{dBB}}^\Psi(\Sigma,\tau)$ distribution. \\
\indent Let's start by looking at the problem in a more general way, and try to answer the two points (i) and (ii) mentioned in the introduction. Since the ideal dBB probability distribution $\mathcal{P}_{\textrm{dBB}}^\Psi(\Sigma,\tau)$ depends on the projected current $|\mathbf{J}^\Psi(\mathbf{x},t)\cdot\mathbf{n}(\mathbf{x})|$, the first question is whether we can associate a POVM with $|\mathbf{J}^\Psi(\mathbf{x},t)\cdot\mathbf{n}(\mathbf{x})|$. The answer is no, and was given by D\"{u}rr and colleagues \cite{Vona2013}. In fact \cite{Vona2013} was not interested in $|\mathbf{J}^\Psi(\mathbf{x},t)\cdot\mathbf{n}(\mathbf{x})|$ but in $\mathbf{J}^\Psi(\mathbf{x},t)\cdot\mathbf{n}(\mathbf{x})$, but the answer is the same.  Let's summarize the reasoning:   
Assume two wave functions $\Psi_1$ and $\Psi_2$ such that $\mathbf{J}^{\Psi_1}(\mathbf{x},t)\cdot\mathbf{n}(\mathbf{x})>0$ and $\mathbf{J}^{\Psi_2}(\mathbf{x},t)\cdot\mathbf{n}(\mathbf{x})>0$ are true at point $\mathbf{x}$ and time $t$, but such that for the wave functions $\Psi_+=\frac{\Psi_1+\Psi_2}{\sqrt{2}}$ and $\Psi_-=\frac{\Psi_1-\Psi_2}{\sqrt{2}}$  we have $\mathbf{J}^{\Psi_+}(\mathbf{x},t)\cdot\mathbf{n}(\mathbf{x})>0$ and $\mathbf{J}^{\Psi_-}(\mathbf{x},t)\cdot\mathbf{n}(\mathbf{x})<0$ (such situations can occur during interference  experiments).   If we assume that $|\mathbf{J}^\Psi(\mathbf{x},t)\cdot\mathbf{n}(\mathbf{x})|$ is a POVM, then by definition we must have:
\begin{eqnarray}
|\mathbf{J}^{\Psi_1}(\mathbf{x},t)\cdot\mathbf{n}(\mathbf{x})|+|\mathbf{J}^{\Psi_2}(\mathbf{x},t)\cdot\mathbf{n}(\mathbf{x})|
=|\mathbf{J}^{\Psi_+}(\mathbf{x},t)\cdot\mathbf{n}(\mathbf{x})|+|\mathbf{J}^{\Psi_-}(\mathbf{x},t)\cdot\mathbf{n}(\mathbf{x})|
\end{eqnarray}
Furthermore, we also have 
\begin{eqnarray}
\mathbf{J}^{\Psi_1}(\mathbf{x},t)\cdot\mathbf{n}(\mathbf{x})+\mathbf{J}^{\Psi_2}(\mathbf{x},t)\cdot\mathbf{n}(\mathbf{x})
=\mathbf{J}^{\Psi_+}(\mathbf{x},t)\cdot\mathbf{n}(\mathbf{x})+\mathbf{J}^{\Psi_-}(\mathbf{x},t)\cdot\mathbf{n}(\mathbf{x})
\end{eqnarray}
Clearly the two relations contradict each other, so $|\mathbf{J}^\Psi(\mathbf{x},t)\cdot\mathbf{n}(\mathbf{x})|$ cannot generally be a POVM. 
This is a central result that rules out any possibility of $\mathcal{P}_{\textrm{dBB}}^\Psi(\Sigma,\tau)$ being a POVM! Notre that this proof was obtain in 2013 \cite{Vona2013} 10 years before the DD  and GTZ results \cite{DasDurr2018,GTZ} and is not depending on spin. This is therefore a very robust result. \\\
\indent But now comes the crux. What is the physical meaning of a POVM $\hat{O}$? Beside mathematics this operator is just a tool, an algorithm,  such that for any wave function $\Psi$ the quantity $\langle \Psi|\hat{O}|\Psi\rangle$ gives us a probability. Physically speaking, it means that we actually have a very precisely defined experimental context or setup (i.e., with external fields, mechanical frames  and so on) such that we can record statistical data proportional to $\langle \Psi|\hat{O}|\Psi\rangle$. The fact that $|\mathbf{J}^\Psi(\mathbf{x},t)\cdot\mathbf{n}(\mathbf{x})|$  is not a POVM implies that there is no experimental configuration such that the amount of statistical information recorded  is directly proportional to $|\mathbf{J}^\Psi(\mathbf{x},t)\cdot\mathbf{n}(\mathbf{x})|$ and this whatever the $\Psi$ wave function chosen. \\
\indent  However,  we should be careful with this theorem: Indeed, this result in no way implies the non-existence of a POVM, or more precisely of an experimental context, which -in some situations and for some $\Psi$- could approximately imply a probability approaching $|\mathbf{J}^\Psi(\mathbf{x},t)\cdot\mathbf{n}(\mathbf{x})|$. Far from that, we all know that detectors are not universal but instead have an optimum operating range outside of which reliable measurement is no longer possible. Therefore, even if it is not possible to build an universal (POVM) detector such that $|\mathbf{J}^\Psi(\mathbf{x},t)\cdot\mathbf{n}(\mathbf{x})|$ is a probability one expect to have POVM detectors such that 
\begin{eqnarray}
\mathcal{P}_{\textrm{detec.}}^\Psi(\mathbf{x},t)= \eta^\Psi|\mathbf{J}^\Psi(\mathbf{x},t)\cdot\mathbf{n}(\mathbf{x})|d\Sigma_\mathbf{x}
\end{eqnarray} where $\eta^\Psi$ is an efficiency coefficient which is in general a complicated function of the quantum state $\Psi$. $\mathcal{P}_{\textrm{detec.}}^\Psi(\mathbf{x},t)$ depends on the detector used and therefore only reproduces approximately the dBB flux predictions given by Eq.~\ref{dBBflux}. In the next sections we will consider the implications of this possibility.
\section{The Fabry-Perot ideal absorbing medium for a plane wave}
\label{section4}
\indent We first consider the non-relativistic problem for spinless particles. We start by supposing a spinless nonrelativistic plane wave $\Psi^{(0)}=e^{ik_1 z}e^{ik_{x}x}e^{-i\frac{k^2}{2m}t}$ impinging on a absorbing Fabry-Perot absorbing slab located between the surfaces $z=0$ and $z=d$ (details concerning this model are given in Appendix~\ref{appendix1}). The number of particles absorbed by the slab intuitively gives us the number of particles detected, and we can apriori define the arrival time distribution as 
\begin{eqnarray}
\mathcal{P}_{\textrm{detec.}}^\Psi(\Sigma,t)=\Sigma J_z^{\Psi^{(0)}}[1-|R|^2-|T|^2]
\end{eqnarray} where $|R|^2$ and $|T|^2$ are the reflection and transmission coefficients respectively (in absence of absorption we would have $1-|R|^2-|T|^2=0$), and where $J_z^{\Psi^{(0)}}=v\cos{\theta}>0$ is the incident current ($v=k/m$ is the de Broglie velocity and $\theta$ the incidence angle with respect to the $z$ axis). We can alternatively write
\begin{eqnarray}
\mathcal{P}_{\textrm{detec.}}^\Psi(\Sigma,t)=-2\Sigma\int_{z=0}^{z=d}dz\textrm{Im}[V_{eff}]|\Psi|^2(x,y,z) \label{detex}\nonumber\\
\end{eqnarray} where $V_{eff}$ is an effective dissipative potential such that $\textrm{Im}[V_{eff}]<0$ (for a discussion of complex potentials in scattering theory see \cite{Schiff}).

\begin{figure}[h]
\includegraphics[width=7cm]{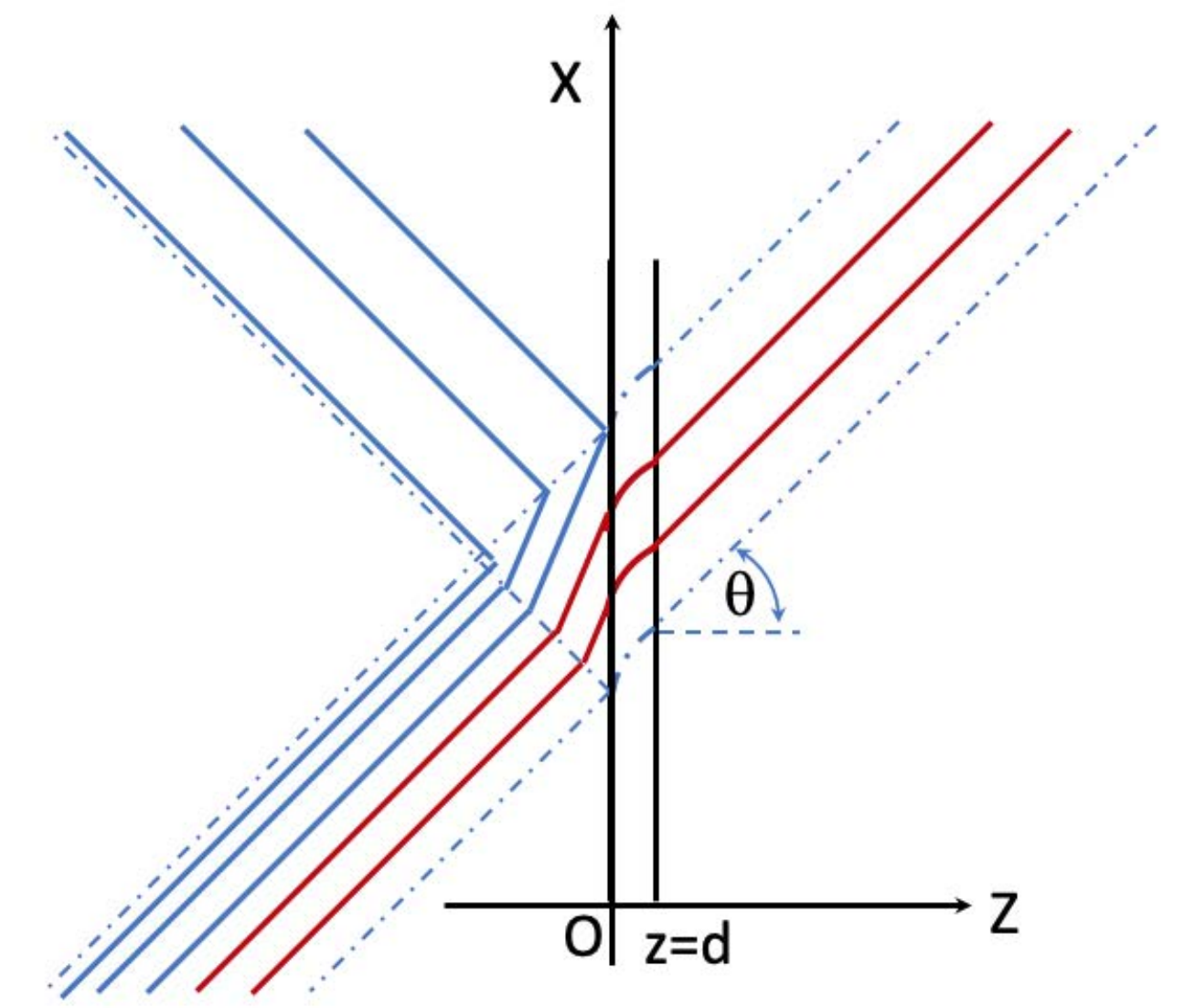} 
\caption{Visualization of the typical dBB trajectories scattered by a thin slab corresponding to a potential barrier.  The dBB trajectories can not cross and therefore the reflected (blue curves) trajectories and transmitted (red curves) trajectories are not overlapping. This is an idealised and schematic representation based on rectilinear rays inspired by the work of Norsen \cite{Norsen}.  In his work Norsen considers a 1D problem and looks at motion in the $t-x$ plane ($t$ being time) whereas here we are looking at a stationary problem in the $x-z$ plane.  } \label{Fig1}
\end{figure}

This complex potential implies a violation of unitarity and the local conservation law is modified as 
\begin{eqnarray}
\partial_t|\Psi|^2=-\boldsymbol{\nabla}\cdot\mathbf{J}^\Psi+2\textrm{Im}[V_{eff}]|\Psi|^2\label{fluxdis}
\end{eqnarray} In this model the violation of unitarity is reminiscent from a coupling with an external bath allowing inelastic scattering through the medium \cite{Halliwell}. Absorbing effective potentials are often used in quantum optics since the 1990's in order to model attenuators and losses. In other words these complex potentials are associated with transmission channels that we can neglect or that we can associate with trapped particles moving outside the domain of propagation considered.     More physically, the medium constituting the slab is filled with absorbing atoms with individual extinction cross-section  $\sigma_{ext}=\sigma_{scat}+\sigma_{abs}$ and we have $-2\textrm{Im}[V_{eff}]=Nv\sigma_{ext}$ where $N$ is the absorbing atom density in the slab. A typical way for justifying such a model is to consider scattering of a particle by a potential well $V(x)$ having one bound energy state $E_0<0$ and a continuum of propagative modes $E_k>0$ coupled to a bath with a continuum of energy levels (plus a vaccum state).  The coupling allows us to derive an   effective potential associated with absorption by the  well, i.e., $V(x)\rightarrow V(x)+ \Delta E-i\Gamma/2$ with $\Gamma>0$ a decay constant associated with dissipation and absorption. \\
\indent Moreover, even in this simple model the physical interpretation must be done carefully and take into account the whole experimental configuration. This is so because dBB trajectories are in general highly contextual and nonclassical. The situation is non ambiguous if the incident wave is actually a `plane wave packet' with finite lateral extension (see Figure \ref{Fig1}) as analyzed for example by Norsen in \cite{Norsen}. Far away from the slab in the $z<0$ domain the incident and reflected contribution are non overlapping and the dBB trajectories are straight lines with constant velocity $v$. In the vicinity of the slab the incident and reflected part interfere and the dBB dynamics is much more complicated. As shown in details in \cite{Norsen} the particle trajectories oscillate around main trajectories sketched in Figure \ref{Fig1} (see also \cite{DewdneyHiley1982,BohmHiley,Holland}). The stream lines separate 2 regions of the initial plane wave packet. The trajectories drawn  in blue are reflected and constitute a fraction $|R|^2$ of the incident flux \cite{Norsen}. The trajectories drawn in red are transmitted through the slab. Moreover because of Eq.~\ref{fluxdis} a fraction of the particle crossing the medium are continuously absorbed  (i.e., detected) along the trajectories. After integrating Eq.~\ref{fluxdis} over a  dBB trajectory $\mathbf{x}(t)$ we get 
\begin{eqnarray}
|\Psi|^2(\mathbf{x(t)},t)=|\Psi|^2(\mathbf{x(t_0)},t_0)e^{-\int_{t_0}^tdt'[\boldsymbol{\nabla}\cdot\mathbf{v}^\Psi_{t'}+2\textrm{Im}[V_{eff}]]}
\end{eqnarray} which contains an exponentially decaying term involving $\textrm{Im}[V_{eff}]<0$. \\
\indent An ideal detector would be such that $|T|^2\simeq 0$ and $|R|^2\simeq 0$. In general this is not so. Ultimately as show in Appendix~\ref{appendix1}, in the weak-coupling regime where absorption is very low the incident flow is weakly disturbed (the medium is nearly transparent $|T|^2\simeq 1$ and $|R|^2\simeq 0$)  and  the arrival-time probability reads  $\mathcal{P}_{\textrm{detec.}}^\Psi(\Sigma,\tau)\simeq  \Sigma d N  \sigma_{ext}|\mathbf{J}^{\Psi^{(0)}}|$ which is independent of the incidence angle $\theta$. In this regime the detector doesn't record the ideal dBB probability  Eq.~\ref{dBBflux}.     Moreover, as shown in Appendix~\ref{appendix1} even if $|T|^2\simeq 0$ it is in general not possible with this simple model to have $|R|^2\simeq 0$. This is not only a question of numerical factor $\eta^\Psi=[1-|R|^2-|T|^2]$ but also an important experimental issue since the reflected and transmitted beams can subsequently disturb and even ultimately prohibit the current flow in other places where different detectors could be located. 
\section{Ideal absorption of a plane wave by a perfectly matched layer}
\label{section5}
\begin{figure}[h]
\includegraphics[width=13cm]{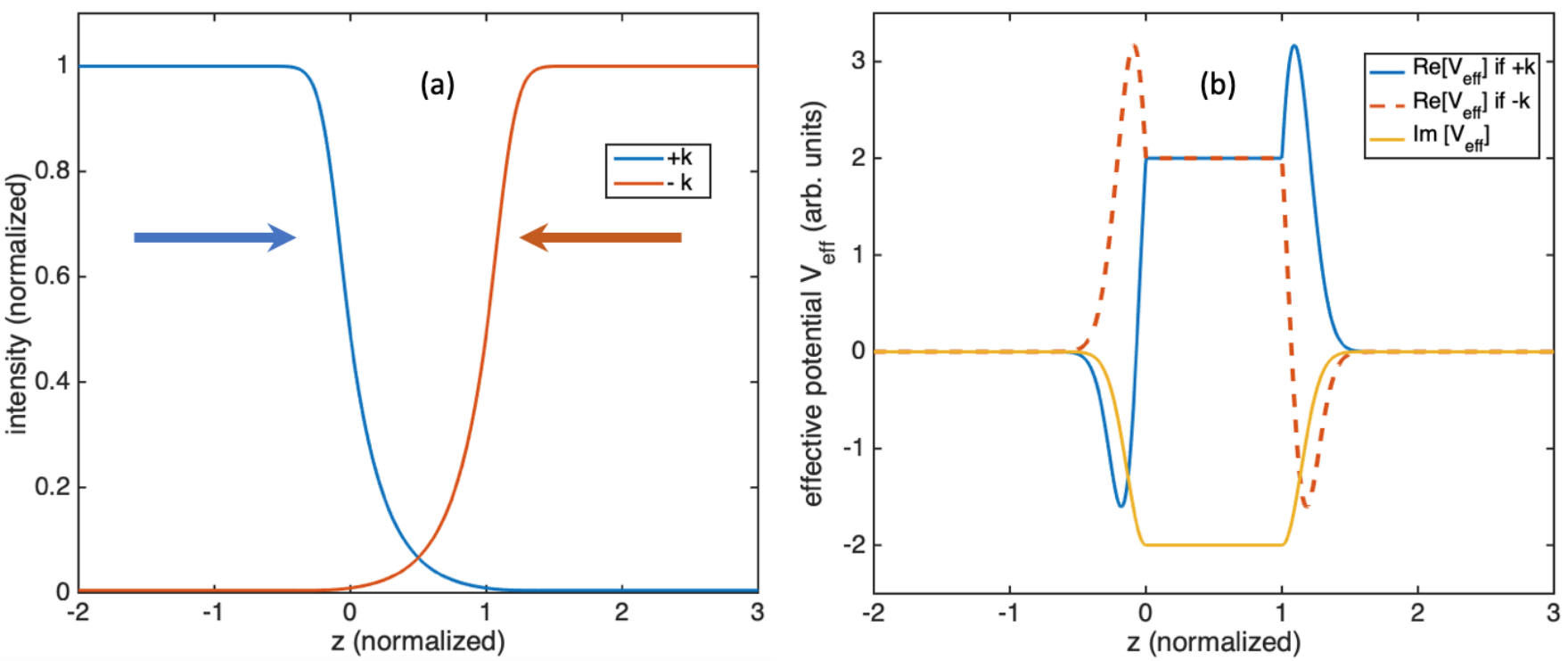} 
\caption{A typical complex potential associated with a PML for a particle detector. In (a) we show the transmitted intensity if the detector is optimized for either a plane wave propagating along the $+z$ direction (blue curve)  or a wave moving along the $-z$ direction (red curve). (b) The real part $\textrm{Re}[V_{eff}]$ of the potentials are shown  (red dashed curve for  the $-z$ incident direction and blue curve for the $+z$ direction) and compared with the imaginary part  $\textrm{Im}[V_{eff}]<0$ of the potential (orange curve) which takes the same form in both $\pm z$ cases.} \label{Fig2}
\end{figure}

\indent The previous model  of Section \ref{section4} was too simple since we considered a homogeneous potential barrier.  The idea to optimize the detector by using stratified media and complex potential has been considered in \cite{Palao1998,Muga1999,Muga1995}. However we here consider a more idealized approach.   Indeed, in principle, an ideal detector can be obtained using a stratified medium known as a perfectly matched layer (PML) often used in numerical calculations \cite{Berenger} (our method differs from the Robin boundary condition approach advocated by Tumulka \cite{Tumulkarobin}). Consider here a one dimensional problem and let $\Psi^{(0)}=e^{ikz}e^{-i\frac{k^2}{2m}t}$  be a plane wave solution of the Schr\"odinger equation with $k=\sqrt{(2mE)}$. We then suppose an ideal  absorbing medium located in the region $z>0$ such that the new wave function  reads   
\begin{eqnarray}
\Psi^{(abs)}(z,t)=e^{ikz}e^{-\int_{-\infty}^zdz'\chi(z')}e^{-i\frac{k^2}{2m}t}\label{trucc}
\end{eqnarray} with $\chi(z)\geq 0$ an absorption function (ideally) vanishing for $z<0$.  As shown in Appendix~\ref{appendix2} we can immediately check that $\Psi^{(abs)}$ is a solution of the equation
\begin{eqnarray}
\partial_z^2\Psi^{(abs)}+2m(E-V_{eff})\Psi^{(abs)}=0
\end{eqnarray}
with the effective complex  potential 
\begin{eqnarray}
V_{eff}(z)=\frac{\chi^2(z)-\chi'(z)}{2m}-i\chi(z)\frac{k}{m}.\label{pot}
\end{eqnarray} corresponding to a dissipative (absorbing) medium or detector. Importantly, for this medium there is no reflected wave (i.e., $R=0$) and the transmitted wave is expoentially decaying as  $|\Psi^{(abs)}(z,t)|^2=e^{-2\int_{-\infty}^zdz'\chi(z')}$ for $z>0$. If the exponentially factor is very large the transmission goes to zero very quickly as required for a good detector. Going back to the detecting slab considered previously we can still apply Eq.~\ref{detex}  for a plane wave at normal incidence if the function $\chi(z)$ ideally vanishes for $z<0$ and $z>d$ (see Figure \ref{Fig2} for a more realistic situation where $\chi(z)$ is a continuous function). We have
\begin{eqnarray}
\mathcal{P}_{\textrm{detec.}}^\Psi(\Sigma,t)=2\Sigma\frac{k}{m}\int_{z=0}^{z=d}dz\chi(z)e^{-2\int_0^zdz\chi(z)}.  \label{detexB}\nonumber\\
\end{eqnarray}
In the particular case where $\chi(z)=\chi_0>0$ is constant for $0<z<d$ (i.e., $\chi(z)=\chi_0\theta(z)\theta(d-z)$) we obtain the effective potential 
\begin{eqnarray}
V_{eff}(z)=\frac{\chi_0^2\theta(z)\theta(d-z)-\chi_0(\delta(z)-\delta(d-z))}{2m}-i\chi_0\theta(z)\theta(d-z)\frac{k}{m}\label{pota1}\end{eqnarray} and Eq.~\ref{detexB} reduces to 
\begin{eqnarray}
\mathcal{P}_{\textrm{detec.}}^\Psi(\Sigma,t)=\Sigma\frac{k}{m}(1-e^{-2\chi_0d})
=\Sigma J_z^{\Psi^{(0)}}(1-e^{-2\chi_0d})  \label{detexC}\nonumber\\
\end{eqnarray} with $J_z^{\Psi^{(0)}}=\frac{k}{m}=v>0$. This detector has an efficiency $\eta^\Psi=1-e^{-2\chi_0d}$. In the limit where $\chi_0 d\rightarrow +\infty$ we have thus $\mathcal{P}_{\textrm{detec.}}^\Psi(\Sigma,t)\rightarrow\Sigma J_z^{\Psi^{(0)}}$ which recovers the dBB arrival time distribution Eq.~\ref{dBBflux}.\\
\indent It must be stressed that the detector is optimized here for a given wavevector $k$ and that in general for a different choice $k\rightarrow k+\delta k$ the potential will not act as a perfect absorber (i.e., in general the reflectivity $R\neq0$ for $\delta k\neq 0$). In a similar way, observe that $k$ and $\chi$ do not necessarily have to be positive and that we can develop a detector adapted to a counterpropagating wave $\propto e^{-ikz}$ with $-k=-\sqrt{(2mE)}<0$ as well. From Eq.~\ref{pot} we still have $\textrm{Im}[V_{eff}]=-\chi_0\theta(z)\theta(d-z)\frac{k}{m}<0$ if $\chi(z)=-\chi_0\theta(z)\theta(d-z)<0$ and this again corresponds to a absorbing medium (the choice $\chi(z)=+\chi_0\theta(z)\theta(d-z)>0$ would have involved  an anti-thermodynamical medium with gain, i.e, emitting particles instead of absorbing them). From Eq.~\ref{detexC} we have now
\begin{eqnarray}
\mathcal{P}_{\textrm{detec.}}^\Psi(\Sigma,t)=\Sigma\frac{k}{m}(e^{2\chi_0d}-1)
=\Sigma |J_z^{\Psi^{(0)}}|(e^{2\chi_0d}-1).  \label{detexD}\nonumber\\
\end{eqnarray} This detector is apriori associated with a different efficiency $\eta^\Psi=e^{2\chi_0d}-1$. However, this is mostly a problem of convention concerning the role of the input and exit sides.   If we instead normalize the field by its value at $z=d$ and not at $z=0$ (this is natural since the wave is counter-propagating and decaying in the $-z$ direction) we recover
$\mathcal{P}_{\textrm{detec.}}^\Psi(\Sigma,t)=\Sigma |{J}_z^{{\Psi'}^{(0)}}|(1-e^{2\chi_0d})$ with the efficiency $\eta^{\Psi'}=1-e^{-2\chi_0d}$ as in Eq.~\ref{detexC} and now ${J}_z^{{\Psi'}^{(0)}}=-\frac{k}{m}e^{-2\chi_0d}$ is associated with the incident plane wave ${\Psi'}^{(0)}=e^{-\chi_0d}e^{-ikz}e^{-i\frac{k^2}{2m}t}$ and Eq.~\ref{trucc} is replaced by 
${\Psi'}^{(abs)}(z,t)=e^{-\chi_0d}e^{ikz}e^{-\int_{-\infty}^zdz'\chi(z')}e^{-i\frac{k^2}{2m}t}$ with $\chi(z)=-\chi_0\theta(z)\theta(d-z)<0$.\\
\indent However, we emphasize that the new effective potential for the back propagating wave is actually different from Eq.~\ref{pota1} since we have 
\begin{eqnarray}
V_{eff}(z)=\frac{\chi_0^2\theta(z)\theta(d-z)+\chi_0(\delta(z)-\delta(d-z))}{2m}-i\chi_0\theta(z)\theta(d-z)\frac{k}{m}\label{pota1b}\end{eqnarray}
the real part of which differs from that deduced from  Eq.~\ref{pota1}. This demonstrates that it is not possible to use the same absorbing  medium for the $e^{ikz}$  and $e^{-ikz}$ cases. If we had (wrongly) used  Eq.~\ref{pota1} for the $e^{-ikz}$ case, we would have obtained an additional contribution in the form of a reflected plane wave proportional to $e^{ikz}$in the $z>d$ domain (i.e., the medium would not act as an idealized absorber for the counterpropagating wave). 
\indent Of course the discussion is  based on a idealized medium and the presence of Dirac distributions in the potential $V_{eff}(z)$ of Eqs.~\ref{pota1}, \ref{pota1b} shows that the  $\chi=const.$ conditions are too strict.   Moreover these pathogical features can be remoded by considering smooth potentials removing the discontinuities. An example is developped in Appendix \ref{appendix3} .  The conclusions we obtained before are however very general (the detector can even be optimized for any plane wave having the associated wavevector components $k_x,k_z$). As illustratted in Figure.~\ref{Fig2} the field transmitted through the medium is strongly attenuated without reflection. Also the potential is characterized by  $\textrm{Im}[V_{eff}]<0$ and as before we require two different potentials optimized either for the $e^{ikz}$ (forward) or $e^{-ikz}$ (backward) cases. 
\section{Generalization for wave packets and time dependent problems  }
\label{section6}
\indent The previous model, based on the interaction of a plane wave with an absorbing medium, can in principle be generalized to the case of a superposition of plane waves forming a wave packet. This is necessary in order to consider the problem associated with backflow. To do this, we will consider a particular case where the problem seems to be treatable with sufficient precision and rigor. 
Let's consider the case where the initial wave function $\Psi^{(0)}(\mathbf{x},t)$, i.e. in the absence of a detector, is developable in Taylor series in the vicinity of a point $\mathbf{x}_0$. More specifically we assume a constant energy $E$, i.e., $\Psi^{(0)}(\mathbf{x},t)\propto e^{-iEt}$ and write $\Psi^{(0)}(\mathbf{x},t)\simeq\Psi^{(0)}(\mathbf{x}_0) e^{-iEt}e^{i\boldsymbol{\nabla}S(\mathbf{x}_0)\cdot(\mathbf{x}-\mathbf{x}_0)+O((\mathbf{x}-\mathbf{x}_0)^2)} $. This is equivalent to assume that the wave function is locally equivalent to a plane wave with an effective wavevector $\mathbf{k}_{eff}(\mathbf{x}_0)=\boldsymbol{\nabla}S(\mathbf{x}_0)$.  As it was shown by Berry \cite{Berry2010} in general such kind of wave packets can easily develop back-flow in the vicinity of a point $\mathbf{x}_0$ . \\
\begin{figure}[h]
\includegraphics[width=14cm]{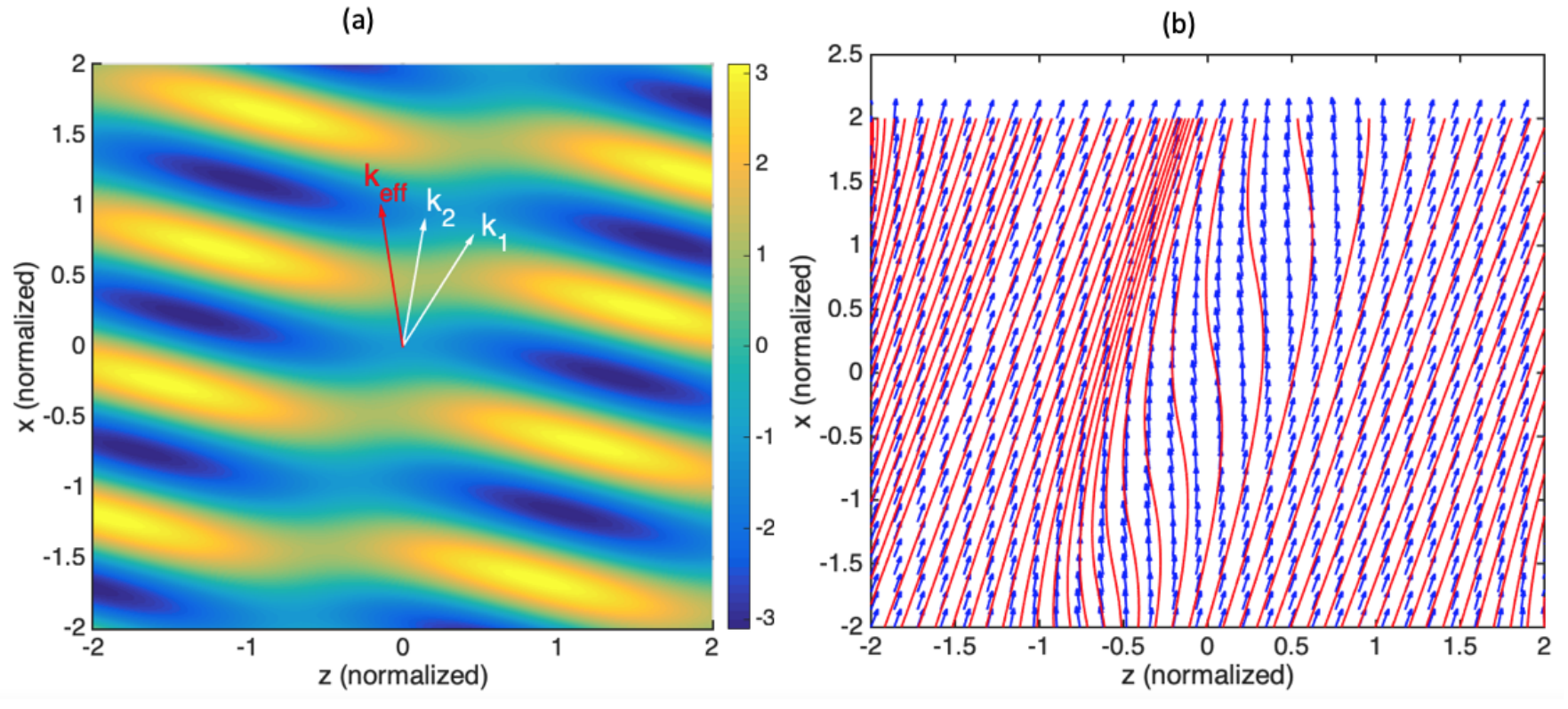} 
\caption{(a) Map of the real part of the wave function $\textrm{Re}[\Psi^{(0)}(z,x)]$ in presence of back-flow (see Eq.~\ref{backflow}) for the case $\mathbf{k}_1=[k_{1z}=k\cos{(\pi/3)},k\sin{(\pi/3)})]$ and $\mathbf{k}_1=[k_{1z}=k\cos{(9\pi/20)},k\sin{(9\pi/20)})]$ and for  $k=2\pi$ (the white arrows show these wave vectors normalized to $k$).  The effective wave vector at the origin $\mathbf{k}_{eff}(0,0)$ (red arrow) has a negative $z-$component due to local back-flow. (b) Map of the dBB  velocity vector and trajectories associated  with map (a).  } \label{Fig3}
\end{figure}
\indent   A simple case is given by the superposition of two plane waves 
\begin{eqnarray}
\Psi^{(0)}(\mathbf{x},t)=(e^{i\mathbf{k}_1\cdot\mathbf{x}}+\alpha e^{i\mathbf{k}_2\cdot\mathbf{x}})e^{-iEt}\label{backflow}
\end{eqnarray}     
with $k_{1z}>0,k_{2z}>0 $ and $\alpha \in \mathbb{C}$ a constant. The probability current reads 
\begin{eqnarray}
\mathbf{J}^{\Psi^{(0)}}(\mathbf{x})=\frac{\mathbf{k}_1}{m}+\frac{\mathbf{k}_1}{m}|\alpha|^2+\frac{\mathbf{k}_1+\mathbf{k}_2}{m}|\alpha|\cos{\Phi}
\end{eqnarray} with $\Phi=(\mathbf{k}_2-\mathbf{k}_1)\cdot\mathbf{x}+\textrm{Arg}[\alpha]$. We are looking for situations where $J^{\Psi^{(0)}}_z<0$ (back flow) and for 	a given $|\alpha|$ we naturally impose $\Phi=\pi$ for points where this backflow is stronger.  Writing  $J^{\Psi^{(0)}}_z=\frac{k_{2z}}{m}f(|\alpha|)$ with 
\begin{eqnarray}
f(|\alpha|)=|\alpha|^2- (1+\frac{k_{1z}}{k_{2z}})|\alpha|+\frac{k_{1z}}{k_{2z}}
\end{eqnarray} we easily find the minimum for $|\alpha|_{min}=\frac{1}{2}(1+\frac{k_{1z}}{k_{2z}})$. In particular, as shown in Appendix  \ref{appendix4},  we easily obtain the effective wavevector $\mathbf{k}_{eff}(\mathbf{x}_0)$ and we have $k_{eff,z}(\mathbf{x}_0=0)=-k_{2z}$ for $\alpha_{min}=-\frac{1}{2}(1+\frac{k_{1z}}{k_{2z}})$.\\ 
\indent It is thus possible to study numerically  the trajectory back flow effect near a point where $\Phi\simeq \pi$ (e.g., near $\mathbf{x}_0=0$). This is illustrated in Figure \ref{Fig3} for a typical example. In this example   we have indeed $k_{eff,z}(\mathbf{x}_0=0)=-k_{2z}$ and we have also $|\mathbf{k}_{eff}(\mathbf{x}_0=0)|\simeq 1.11|k_{2}|$ meaning that the effective wavelength $\lambda_{eff}\simeq 0.9 \lambda_0 $ is very close from the initial values of each plane waves.\\
\indent As shown in Figure \ref{Fig3} the interference zone where the back flow is visible is sufficiently extended to imagine a detector localized in the region $\mathbf{x}_0\simeq 0$ and able to observe the phenomenon. In order to be more precise we can use the model or perfectly matched layer detectors and introduce a local potential barrier $V_{eff}(z)$ (see Eq.~\ref{pota1} or Eq.~\ref{pota1b}) adapted to any regions of the interference field $\Psi^{(0)}(\mathbf{x})$ and this for both the normal and back flow regimes. \\
\indent  The great specificity of these detectors is of course that they are highly optimized for well-defined regions of space (i.e., in the near environment of particular points $\mathbf{x}_0$, $\mathbf{x}_1$, ...) where the initial wave function $\Psi^{(0)}(\mathbf{x},t)$ moves. Obviously, these detectors are highly invasive in the sense that they are optimized to cancel local reflectivity at the chosen point $\mathbf{x}_0$, $\mathbf{x}_1$, ... for a given local effective wave vector $\mathbf{k}_{eff}(\mathbf{x}_0), \mathbf{k}_{eff}(\mathbf{x}_1),...$ Transmission is also zero, which implies high local absorption associated with high detection efficiency $\eta^{\Psi^{(0)}}(\mathbf{x}_0),\eta^{\Psi^{(0)}}(\mathbf{x}_1),...$ \\
\indent Moreover, faraway from  the detector the wave function is in general disturbed due to diffraction and scattering by the potential $V_{eff}$. Indeed, consider a single detector centered on point $\mathbf{x}_0:=[x_0,y_0,z_0]$. We assume that the detector has a finite volume $\delta V$ located between the planes $z=z_0$ and $z_0+d$ where $d$ can be arbitrarily small if the detector is very efficient (due to a fast decay of the wave propagating in the absorbing medium). The transverse extension of the medium in the $x$ and $y$ direction is also limited  around $x=x_0$ and $y=y_0$ (i.e., ideally over few wavelengths $\lambda_{eff}(\mathbf{x}_0)$).  Therefore, the wave function in the vicinity of point $\mathbf{x}_0$ is at a first approximation given by the previous theory.\\
\indent However,  due to the finite extension of the detector important deviations must occur in the far-field.   If we write $K_E^{(0)}(\mathbf{x}|\mathbf{x}_1)$ the  time independent Green's function for the Schr\"{o}dinger equation  in vacuum we have
\begin{eqnarray}
EK_E^{(0)}(\mathbf{x}|\mathbf{x}_1)=\frac{-\boldsymbol{\nabla}^2}{2m}K_E^{(0)}(\mathbf{x}|\mathbf{x}_1)+\delta^3(\mathbf{x}-\mathbf{x}_1)
\end{eqnarray}   and the usual retarded solution is  
\begin{eqnarray}
K_E^{(0)}(\mathbf{x}|\mathbf{x}_1)=-2m\frac{e^{i\sqrt{(2mE)}|\mathbf{x}-\mathbf{x}_1|}}{4\pi|\mathbf{x}-\mathbf{x}_1|}
\end{eqnarray}
From the Green theorem the wave function scattered by the effective potential $V_{eff}$ (with a negative imaginary part: $\textrm{Im}[V_{eff}]<0$) is given by 
\begin{eqnarray}
\Psi(\mathbf{x})=\Psi^{(0)}(\mathbf{x})+\int_{\delta V} d^{3}\mathbf{x}_1K_E^{(0)}(\mathbf{x}|\mathbf{x}_1)V_{eff}(\mathbf{x}_1)\Psi(\mathbf{x}_1)
\end{eqnarray} where the integration is taken over the finite volume  $\delta V$ of the detector i.e., for points $\mathbf{x}_1$ surrounding $\mathbf{x}_0$.\\
\indent The scattered field $\Psi_{scat}(\mathbf{x})=\Psi(\mathbf{x})-\Psi^{(0)}(\mathbf{x})$ is equivalently writen using a surface integral over the closed boundary
$\delta \Sigma$ surrounding  $\delta V$.  From the Huygens-Fresnel theory  applied to Schr\"odinger's equation we deduce:
\begin{eqnarray}
\Psi_{scat}(\mathbf{x})=-\oint_{\delta\Sigma} d\Sigma_1\mathbf{n}_1\cdot[\Psi(\mathbf{x}_1)\boldsymbol{\nabla}_{\mathbf{x}_1}\frac{K_E^{(0)}(\mathbf{x}|\mathbf{x}_1)}{2m}-\frac{K_E^{(0)}(\mathbf{x}|\mathbf{x}_1)}{2m}\boldsymbol{\nabla}_{\mathbf{x}_1}\Psi(\mathbf{x}_1)]
\end{eqnarray}  where $\mathbf{n}_1$ is the outwardly oriented unit vector normal to the surface element  $d\Sigma_1$ at point $\mathbf{x}_1\in \delta\Sigma$.
In the case above of a sensing volume lying between planes $z=z_0$ and $z_0+d$ and assuming a current $J_z^{\Psi^{(0)}}(\mathbf{x}_0)>0$, the integration surface reduces approximately to the input face $\delta\Sigma_{in}$ at $z=z_0$ and we have 
\begin{eqnarray}
\Psi_{scat}(\mathbf{x})\simeq-\int_{\delta\Sigma_{in}} dx_1dy_1[\Psi(\mathbf{x}_1)\partial_{z_1}\frac{K_E^{(0)}(\mathbf{x}|\mathbf{x}_1)}{2m}-\frac{K_E^{(0)}(\mathbf{x}|\mathbf{x}_1)}{2m}\partial_{z_1}\Psi(\mathbf{x}_1)]\label{fresnel1}
\end{eqnarray} with $R=|\mathbf{x}-\mathbf{x}_1|$. In general this scattered field doesn't vanish and in order to get a converging expression we replace the surface integral at $z_0$ by a plane located at $z_0-\epsilon$ such that $\chi(z_0-\epsilon)\simeq 0$. Eq.~\ref{fresnel1} reads thus 
\begin{eqnarray}
\Psi_{scat}(\mathbf{x})\simeq\Psi^{(0)}(\mathbf{x}_0)\int_{\delta\Sigma_{in}} dx_1dy_1\frac{e^{ikR}}{4\pi R}(ik\hat{\mathbf{R}}\cdot\hat{\mathbf{z}}+ik_z)
e^{i\mathbf{k}_{||}\cdot (\mathbf{x}_1-\mathbf{x}_0)}\label{fresnel2}
\end{eqnarray} with $\mathbf{R}=\mathbf{x}-\mathbf{x}_1$, $\hat{\mathbf{R}}=\mathbf{R}/|\mathbf{R}|$ and the vector $k_z\hat{\mathbf{z}}+\mathbf{k}_{||}:=\boldsymbol{\nabla}S(\mathbf{x}_0)$ as explained before (for $k_z>0$).    For points in the shadow region, near the detector, the scattered contribution $\Psi_{scat}(\mathbf{x})$ strongly compensates the incident term $\Psi^{(0)}(\mathbf{x})$ and the full wave function approximately vanishes. However, in general this implies that the scattered wave interfere with the incident one and this will disturb the probability current $\mathbf{J}^{\Psi^{(0)}}$ as well as the dBB trajectories in the vicinity of the strongly absorbing detector located at $\mathbf{x}_0$. The detector is thus invasive and the disturbed trajectory flow will in general prohibit a subsequent measurement of the incident current $\mathbf{J}^{\Psi^{(0)}}$ at a different point $\mathbf{x}'_0$ located near $\mathbf{x}_0$. We stress Eq.~\ref{fresnel2} must be multiplied by the coeeficient $-1$ if the current is counterprogating, i.e., if we  have a local backflow with $k_z<0$ (this is because in this regime the input face at $z=z_0$ contributing to the integral is replaced by the output face at $z=z_0+d$).  We have thus:
\begin{eqnarray}
\Psi_{scat}(\mathbf{x})\simeq-\Psi^{(0)}(\mathbf{x}_0)\int_{\delta\Sigma_{in}} dx_1dy_1\frac{e^{ikR}}{4\pi R}(ik\hat{\mathbf{R}}\cdot\hat{\mathbf{z}}+ik_z)
e^{i\mathbf{k}_{||}\cdot (\mathbf{x}_1-\mathbf{x}_0)}\label{fresnel3}
\end{eqnarray} and all the  conclusions concerning the invasiveness of the detector are of course also valid in this backflow regime.\\
\indent The previous  difficulties concerning ideal detectors are very general and will apply to current measurements for time dependent problems. 
This in principle central for time of flight and arrival time measurements. Qualitatively the problem involves  absorbing detectors modeled by time dependent 
dissipative potentials $V_{eff}(\mathbf{x}, t)$. The potential is supposedly acting only in a small regions of space $\delta V$ surrounding a point $\mathbf{x}_0$ during a 
time interval $\delta t$ surrounding a time $t_0$.\\ 
\indent  The central formula in the above (non-relativistic) analysis is Eq.~\ref{fluxdis}  
\begin{eqnarray}
-\partial_t|\Psi|^2=\boldsymbol{\nabla}\cdot\mathbf{J}^\Psi-2\textrm{Im}[V_{eff}]|\Psi|^2\label{fluxdisb}
\end{eqnarray} in which the sink term $-2\textrm{Im}[V_{eff}]|\Psi|^2\geq 0$ represents the local absorption of the medium characterized by an effective dissipative potential with $\textrm{Im}[V_{eff}]<0$. The probability of absorbing a particle is thus generally given by
\begin{eqnarray}
\mathcal{P}_{\textrm{detec.}}^\Psi(\delta \Omega)=-2\int_{\delta\Omega}d^4x\textrm{Im}[V_{eff}(\mathbf{x},t)]|\Psi|^2(\mathbf{x},t) \label{detex4D}
\end{eqnarray} where $\delta \Omega$ is a 4-volume in space-time where the detector is active  and the effective potential $V_{eff}(\mathbf{x},t)\neq 0$. This effective potential is associated with relaxation and dissipation and can ultimately be justified by interactions with a thermal bath (see Section \ref{section4}).  We stress that $\mathcal{P}_{\textrm{detec.}}^\Psi(\delta \Omega)$ is a POVM since since $-\textrm{Im}[V_{eff}(\mathbf{x},t)]\geq 0$ and $|\Psi|^2(\mathbf{x},t)=\langle\mathbf{x}|\Psi_t|\mathbf{x}\rangle$ (we have the additivity for two disjoints regions $\delta \Omega_1$ and  $\delta \Omega_2$ $\mathcal{P}_{\textrm{detec.}}^\Psi(\delta \Omega_1\cup \delta \Omega_2)=\mathcal{P}_{\textrm{detec.}}^\Psi(\delta \Omega_1)+\mathcal{P}_{\textrm{detec.}}^\Psi(\delta \Omega_2)$). The measurement of the probability $\mathcal{P}_{\textrm{detec.}}^\Psi(\delta \Omega)$ is thus physically unambiguous and must agree in both the orthodox and Bohmian quantum interpretations. \\
\indent In practice, however, it is extremely difficult to build a detector with space-time resolution. The basic idea, for example, would be to introduce a time-shutter that opens and closes in a narrow time window $\delta t$. Within this time  window, the incident particle is likely to pass through and interact with the absorbing medium of the detector. However, the wave theory of time-shutters and transient phenomena linked to diffraction in time  is complex (see, for example \cite{Moshinsky1952,Kleber1994,Brukner1997}) and we won't go into it here. Calculations not reproduced here show in particular that the presence of the shutter strongly disturbs the incident wave field (e.g.  due to the presence of back-scattering), and this will of course have an impact on the $\mathcal{P}_{\textrm{detec.}}^\Psi(\delta \Omega)$ probability. Another method could be to use a dynamical potential barrier \cite{Park}, or alternatively a metal plate that rejects secondary electrons when subjected to a local excitation in space-time  \cite{Mlynek1997} (this approach has been used in an interferometry experiment involving He atoms \cite{Kurtsiefer1997} and analyzed using dBB dynamics \cite{Das2022}). For a review of detection methods relevant for arrival time measurement see \cite{Muga2000,Dassreview} as well as \cite{Wolf,Stopp,Steinhauer}.  \\
\indent Limiting our description to an effective absorption potential and Eq.~\ref{detex4D} it will be in general difficult to reduce the probability to the simple dBB formula Eq.~\ref{dBBflux}. Going back to Eq.~\ref{fluxdisb} and integrating over a four-volume $\delta \Omega=\delta V \times\delta t$ we get by applying Gauss theorem:
\begin{eqnarray}
\int_{\delta V}d^3x |\Psi(\mathbf{x},t)|^2-\int_{\delta V}d^3x |\Psi(\mathbf{x},t+\delta t)|^2 \nonumber\\ +\int_t^{t+\delta t} dt\oint_\Sigma d^2\Sigma_\mathbf{x}\cdot\mathbf{n}_\mathbf{x} \mathbf{J}^\Psi(\mathbf{x},t) =-2\int_{\Omega}d^4x\textrm{Im}[V_{eff}(\mathbf{x},t)]|\Psi|^2(\mathbf{x},t) 
\end{eqnarray}  where $\mathbf{n}_\mathbf{x}$ is the inward oriented unit vector normal to the closed boundary $\Sigma$ surrounding the volume $\delta V$ of the detector. If the detector is efficient we naturally expect $\int_{\delta V}d^3x |\Psi(\mathbf{x},t+\delta t)|^2 \simeq 0$. Similarly for an efficient and compact detector reducing to a slab we must have $\int_{\delta V}d^3x |\Psi(\mathbf{x},t)|^2\ll \int_t^{t+\delta t} dt\oint_\Sigma d^2\Sigma_\mathbf{x}\mathbf{n}_\mathbf{x} \cdot\mathbf{J}^\Psi(\mathbf{x},t)\simeq  \delta t\int_{\Sigma_{in}} d^2\Sigma_\mathbf{x}\mathbf{n}_\mathbf{x} \cdot\mathbf{J}^\Psi(\mathbf{x},t)$ where the only important contribution of the surface integral comes from the entrance surface of the detector $\Sigma_{in}$ (the sign $\int_{\Sigma_{in}} d^2\Sigma_\mathbf{x}\mathbf{n}_\mathbf{x} \cdot\mathbf{J}^\Psi\geq 0$ is thus naturally imposed but we can introduce a minus sign if we need to consider a back flow process). Therefore Eq.~\ref{detex4D} reduces to  
\begin{eqnarray}
\mathcal{P}_{\textrm{detec.}}^\Psi(\delta \Omega)=-2\int_{\Omega}d^4x\textrm{Im}[V_{eff}(\mathbf{x},t)]|\Psi|^2(\mathbf{x},t)
\simeq \delta t\int_{\Sigma_{in}} d^2\Sigma_\mathbf{x}\mathbf{n}_\mathbf{x} \cdot\mathbf{J}^\Psi(\mathbf{x},t)
\label{detex4Dbis}
\end{eqnarray} which is recovering  the dBB  flux result Eq.~\ref{dBBflux} with a detecting efficiency $\eta \simeq 1$. Of course, the present analysis is only an approximation. It cannot be general since by definition  Eq.~\ref{detex4D}  is a POVM whereas from the theorem of D\"{u}rr et al. \cite{Vona2013} (derived in Section \ref{section3})  $\mathcal{P}_{\textrm{dBB}}^{\Psi}(\mathbf{x},\tau)$ is not a POVM!  We insist on the fact that if we use the perfectly matched detector layer studied previously, then the surface integral $\int_{\Sigma_{in}} d^2\Sigma_\mathbf{x}\cdot\mathbf{n}_\mathbf{x} \mathbf{J}^\Psi(\mathbf{x},t)$ depends on the initial wave function $\Psi^{(0)}$ existing in the absence of a detector. In this case, we were justifying the possibility of measuring the local dBB distribution. In other words, we have the local equivalence (and for this wave function): $\mathcal{P}_{\textrm{detec.}}^\Psi(\delta \Omega)\simeq \delta t \mathcal{P}_{\textrm{dBB}}^{\Psi^{(0)}}(\mathbf{x},\tau)$. \\
\indent Furthermore, like in the stationnary regime, the far-field wave function will be in general strongly modified since we will have 
\begin{eqnarray}
\Psi(\mathbf{x},t)=\Psi^{(0)}(\mathbf{x},t)+\int_{\delta \Omega} d^{4}x_1 K^{(0)}(\mathbf{x},t|\mathbf{x}_1,t_1)V_{eff}(\mathbf{x}_1,t_1)\Psi(\mathbf{x}_1,t_1)
\end{eqnarray}  where $K^{(0)}(\mathbf{x},t|\mathbf{x}_1,t_1)= -i(\frac{m}{2\pi i(t-t_1)})^{3/2}e^{i\frac{m(\mathbf{x}-\mathbf{x}_1)^2}{2(t-t_1)}} \Theta(t-t_1)$ is the retarded Schr\"{o}dinger Green function for the time-dependent problem. This implies that the mere presence of an efficient absorbing detector in the space-time region $\delta \Omega$ will in general disturb and influence the surrounding environement. In particular it will generally affects other detectors as we will now see.   
\section{General discussions and conclusions: Can we observe Bohmian first arrival time ? (second round)}
\label{section7}
\subsection{Weak coupling versus strong coupling: Advantages and limitations}
\begin{figure}[h]
\includegraphics[width=9cm]{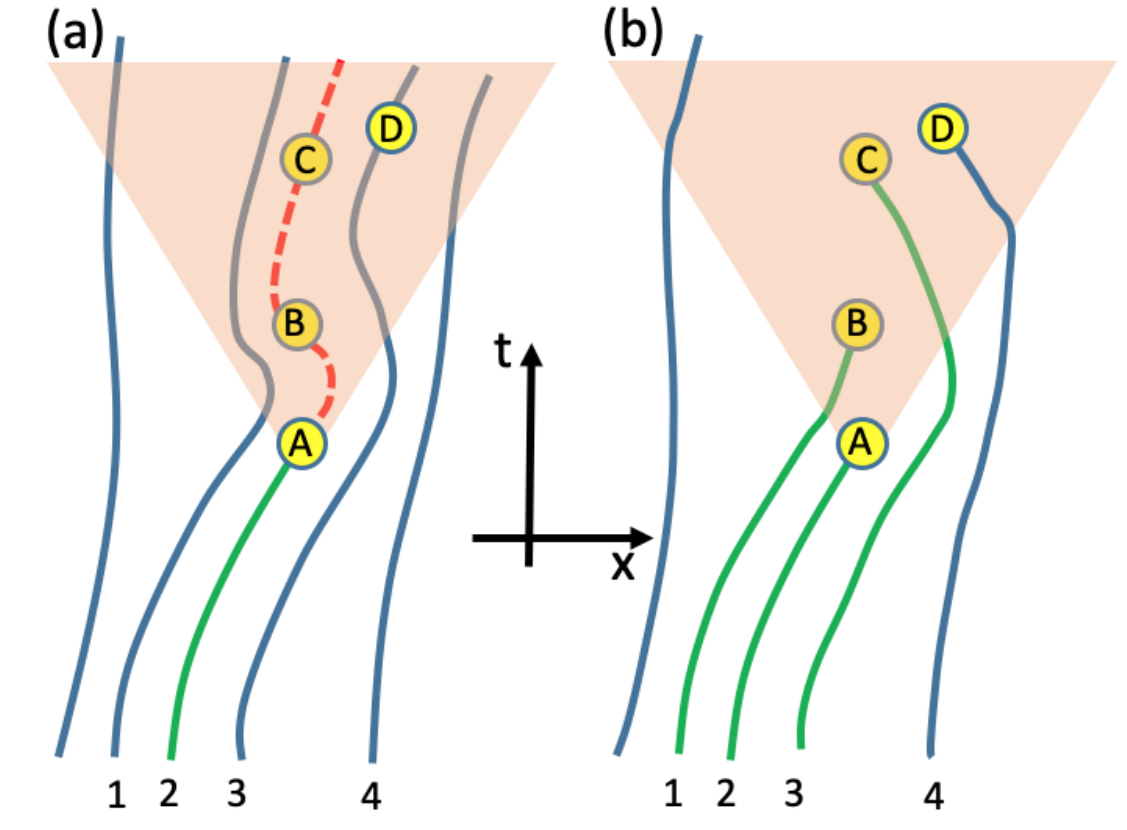} 
\caption{Bohmian trajectories in presence of position-time detectors. (a) In the idealized weak-coupling regime detectors at points A, B, C or D are not perturbed by spurious scattering and the system is sensitive to the incident wave function $\Psi^{(0)}$ existing in the absence of detectors.  However one must add a clock to distinguish precisely   the first arrival, second arrival etc... at a given position. This requires a precise knowledge od the dBB trajectories. (b) In the strong regime only one pass is needed: The detectors absorbe with  high efficiency the incoming particles.  However, this strongly affects the wave function $\Psi \neq \Psi^{(0)}$ and disturbs the dBB motion on other detectors.     } \label{Fig4}
\end{figure}

\indent The previous results show that, in general, it should not be impossible to measure the dBB probability distribution $\mathcal{P}_{dBB}^\Psi(\mathbf{x},t) :=|\mathbf{J}^\Psi(\mathbf{x},t)\cdot\mathbf{n}(\mathbf{x})|d\Sigma_\mathbf{x}$ of the first arrival times at a given space-time point. However, as we show now the procedure is generally very invasive and may prevent this distribution from being measured in several space-time regions in the same experiment. This point is crucial and is in line with the result obtained on the no-signalling theorem concerning the impossibility of measurements that would contradict special relativity.\\
\indent Indeed, the central problem we have seen in previous sections is that an effective detector modifies the $\Psi$ wavefield in its immediate environment by scattering. This is in some ways reminiscent of Renninger's results on null or negative measurements: A non-measurement or non-detection of a particle by a localised screen impacts the wave function of the particle outside that screen. The effect can never be neglected specifically if the wavelength of $\Psi$ is comparable to the detector size.\\ 
\indent In Bohmian mechanics the consequences are unavoidable in order to understand how  the presence of detectors can affect subsequent potential interactions or detections. The problem is fundamentally linked to relativistic causality. For example, consider a quantum particle following a Bohmian trajectory in space-time, as shown in Figure \ref{Fig4} (a). Along this trajectory we can place absorbing detectors at the space-time points $A$, $B$, and $C$ (with time $t_C>t_B>tA$). From an intuitive point of view, if the detector observes a particle at $A$, this naturally prevents subsequent observations at $B$ and $C$. In other words, intuitively, if in an experiment we position three detectors at $A$, $B$ and $C$, only the first detector, which is assumed to be very efficient, will be able to potentially observe a particle (this probability of observation being given by Born's rule $|\Psi|^2$ for this point $A$) and we will never observe particles at $B$ or $C$ because the mere presence of the detector at A screens out the other detectors and precludes interactions. This intuitive description is however approximative and essentially classical.  It presupposes that the presence of the detectors at $A$, $B$, $C$ is not influencing the incident wave function $\Psi$. However what we saw in previous sections of this article is precisely the opposite:  The detectors generally disturb the wave function $\Psi$.\\
\indent As we saw we can basically distinguish two regimes. In the `weak coupling regime' the detector is highly inefficient and  $\eta\ll 1$. In this regime the Bohmian trajectories can be considered as approximately non modified: Most of the particles going through the region $A$ of the previous example will not be absorbed by the detector and only a little fraction of the incident particles will contribute to the recording signal at $A$. Moreover, in this weak coupling regime nothing prohibits the detectors at position $B$ or $C$ to fire if the particle has not been detected at $A$ (and $B$ if we consider detection at $C$). Since we can ultimatelly suppose that the incident wave function is not disturbed (i.e., we can neglect scattering in Eq.~\ref{fresnel2}) this implies that at the lowest order of approximation the probability  of detection $\mathcal{P}_{\textrm{detec.}}^{\Psi^{(0)}}(\mathbf{x},t)$ at $A$, $B$ or $C$ are just calculated by ignoring  the presence of the other detectors and using the incident wave function $\Psi^{(0)}$.\\          
\indent There are clearly advantages and disadvantages to considering the weak coupling regime. Starting with the advantages we can see, by returning to Figure \ref{Fig4} (a), that in this regime we can define detection experiments involving several points $A$, $B$, $C$, etc. and associated with complex geometries. From this point of view, measuring the probability distribution $\mathcal{P}_{\textrm{detec.}}^{\Psi^{(0)}}(\Sigma,t)$ is not in principle a problem. We can imagine, for example, a set of weakly absorbing detectors distributed in a finite region of space-time in order to have access in the same experiment to the probability distribution of arrival times associated with the initial wave function $\Psi^{(0)}$.\\
\indent However, this weak coupling regime leads to two important problems.  Firstly, as we have seen in the previous sections, in the non-relativistic regime, if the efficiency of the detector decreases, the  arrival-time probability approaches $\mathcal{P}_{\textrm{detec.}}^\Psi(\Sigma,\tau)\simeq  \Sigma d N  \sigma_{ext}|\mathbf{J}^{\Psi^{(0)}}|$ which is independent of the incidence angle $\theta$ and which  shows that in the non-relativistic regime in which the calculations are carried out, the detector is insensitive to the direction of the probability current. Going back to Eq.~\ref{detex4D} we have for an indiivudal detector  centered on $A$:
\begin{eqnarray}
\mathcal{P}_{\textrm{detec.}}^\Psi(\delta \Omega)\simeq -2\delta t \int_{\delta V}d^3x\textrm{Im}[V_{eff}(\mathbf{x},t)]|\Psi^{(0)}|^2(\mathbf{x},t) \label{detex4Dtri} \sim \epsilon \delta t\delta V |\Psi^{(0)}|^2(\mathbf{x}_A,t_A)\label{badone}
\end{eqnarray} with $\epsilon \sim -\frac{2}{\delta V} \int_{\delta V}d^3x\textrm{Im}[V_{eff}(\mathbf{x},t)] $ a characteristic rate (in the model used previously, see Section \ref{section4},  we have $\epsilon = -2\textrm{Im}[V_{eff}]=Nv\sigma_{ext}$ and $\delta V=\Sigma d$ which allows us to recover the formula $\mathcal{P}_{\textrm{detec.}}^\Psi(\Sigma,\tau)\simeq  \Sigma d N  \sigma_{ext}|\mathbf{J}^{\Psi^{(0)}}|$).
The detection probability is therefore no longer simply related to the ideal Bohmian probability Eq.~\ref{dBBflux} $\mathcal{P}_{dBB}^\Psi(\mathbf{x},t) :=|\mathbf{J}^\Psi(\mathbf{x},t)\cdot\mathbf{n}(\mathbf{x})|d\Sigma_\mathbf{x}$ associated with the probability current $\mathbf{J}^\Psi(\mathbf{x},t)$. \\
\indent The second problem with weak coupling arises from the very fact that the dBB trajectories obtained are unperturbed with respect to the initial wave function $\Psi^{(0)}$ (see Figure \ref{Fig4} (a)). It is therefore possible to imagine a detector that is sensitive to the first pass (at A), the second pass (at B), the third pass (at C), etc. However, if the detector can observe a particle corresponding to the second pass at B, for example (which may be associated with a back-flow phenomenon), a clock is needed that can determine when the particles arrive and tell when it is a first, second or third pass. In practice, this requires knowledge of the dBB trajectories and is therefore dependent on the initial wave function. This problem seems to be even more fundamental than the first, as it concerns the very notion of measurement. In quantum measurement theory, which is based on the notion of POVM, it is presupposed that any good measurement requires detection equipment that will function independently of the chosen initial wave function $\Psi^{(0)}$. This is what is implied by the POVM formalism that reduces any observable probability to an expression of the type $\mathcal{P}^{\Psi}_{a}=\langle \Psi^{(0)}|\hat{O}_{a}|\Psi^{(0)}\rangle$, which contains an operator $\hat{O}_{a}$ (independent of $|\Psi^{(0)}\rangle$) and the wave function itself $|\Psi^{(0)}\rangle$, which averages the operator $\hat{O}_{a}$. Here however in order to define the first, second etc... arrival times the precise knowledge of the dBB trajectories is needed. This clearly seems to violate the natural formulation of quantum measurement based on POVM and therefore doesn't look appealing. For the reasons mentioned above, the weak coupling regime seems apriori unsuitable. This analysis confirms some of the worries of GTZ \cite{GTZ,GTZ2}.   We will see below that, far from being the case, the alleged defects will in fact turn out to be advantageous. However, ignoring  temporarily this point, it seems natural at this stage to focus on the other regime, i.e. the strong coupling regime with highly efficient detectors.  \\
\indent In the strong coupling regime the  detection efficiency is high, i.e.,  $\eta\sim 1$.  As we have seen in this regime, the probability of detection approaches the ideal Bohmian  formula Eq.~\ref{dBBflux} $\mathcal{P}_{dBB}^\Psi(\mathbf{x},t) :=|\mathbf{J}^\Psi(\mathbf{x},t)\cdot\mathbf{n}(\mathbf{x})|d\Sigma_\mathbf{x}$. Moreover, as the probability of absorption is high, this seems de facto to prohibit the detection of second, third ... passes: in principle, only the first pass could be measured. This therefore seems intuitively desirable for a procedure for measuring the first arrival time distribution of a particle in a given zone of space (in agreement with the idealized classical picture of an absorption).\\
\indent Alas, this image is, of course, an oversimplification because, as we have analysed above, the simple registration of a strongly absorbing detector in $A$ will disturb the detector's immediate environment by scattering. Thus, in general, the wave function is locally modified and the dBB trajectories are strongly perturbed.  In the situation shown in Figure \ref{Fig4} (a), the Bohmian trajectories are causally disturbed by diffraction in the future light-cone emerging from point $A$.  In fact, even if the particle passing through A is well absorbed, this in no way precludes particle detection at $B$ and $C$, since other disturbed trajectories may reach regions $B$, $C$, etc., as shown in Figure \ref{Fig4} (b).  Other regions may also be affected, such as the one where detector $D$ is located, which was not on the initial dBB trajectory through ABC!\\
\indent We can obtain a theorem concerning this issue. Indeed, from \cite{Vona2013} and results discussed in Sections \ref{section3},\ref{section6} (in particular Eq.~\ref{detex4Dbis}), we know that $\mathcal{P}_{\textrm{detec.}}^\Psi(\delta \Omega)\simeq \delta t \mathcal{P}_{\textrm{dBB}}^{\Psi^{(0)}}(\mathbf{x},\tau)$ can not be generally true for every wave functions $\Psi$ since the left hand side of the relation is a POVM whereas the right hand side is not. Let's assume that  $\mathcal{P}_{\textrm{detec.}}^\Psi(\delta \Omega)\simeq \delta t \mathcal{P}_{\textrm{dBB}}^{\Psi^{(0)}}(\mathbf{x},\tau)$ is (approximately) true for a specific wave function $\Psi$. We can thus consider several effecient detectors in regions $\delta \Omega_1$, $\delta \Omega_2$,... and suppose that  for a given wave function $\Psi$ we have 
\begin{eqnarray}
\mathcal{P}_{\textrm{detec.}}^\Psi(\cup_i\delta \Omega_i)=\sum_i\mathcal{P}_{\textrm{detec.}}^\Psi(\delta \Omega_i)\simeq  \sum_i \delta t_i \mathcal{P}_{\textrm{dBB}}^{\Psi_i}(\mathbf{x}_i,\tau_i)\label{assumption}
\end{eqnarray}  Here $\Psi_i$ is the local wave function in the region $\delta \Omega_i$ that in principle includes the scattering contributions from other detectors that could causally interact with detector $i$ (i.e., located in its past light-cone). Physically we know some cases where Eq.~\ref{assumption} is certainly true or a very good approximation (e.g., typically in the far-field domain \cite{Mlynek1997,Das2022}). But what we would ideally like to obtain is the stronger result 
\begin{eqnarray}
\mathcal{P}_{\textrm{detec.}}^\Psi(\cup_i\delta \Omega_i)=\sum_i\mathcal{P}_{\textrm{detec.}}^\Psi(\delta \Omega_i)\simeq  \sum_i \delta t_i \mathcal{P}_{\textrm{dBB}}^{\Psi^{(0)}_i}(\mathbf{x}_i,\tau_i)\label{assumptionB}
\end{eqnarray}
which depends on the initial wave function $\Psi^{(0)}_i:=\Psi^{(0)}(\mathbf{x}_i,t_1)$ unperturbed by the presence of detectors and calculated at the various space-time points where the detectors are actually located. For a single isolated detector we apriori know that this is possible but for a set of detector the question remains open. However, we can easily show that this is actually impossible. Indeed, since the various detectors in regions $\delta \Omega_1$, $\delta \Omega_2$,... are strongly efficient they actually record the probability of first arrival in these regions. The situation is thus the one sketched in Figure \ref{Fig4} (b). But   the points $A, B , C$ on this example  are located on the same initial unperturbed trajectory. Therefore, in order to have Eq.~\ref{assumptionB} true, we would have to have detectors located at $B$ and $C$ which are sensitive to the first passage of the particle at this point (in accordance with the definition and Figure \ref{Fig4} (b)) and yet according to Eq.~\ref{assumptionB} in reality these detectors would measure the second passage (for $B$) and the third passage (for $C$). Another way of saying this is that, in accordance with the definition of the strong coupling regime, the detector in $A$ should be able to observe a particle, while those in $B$ and $C$ should observe nothing in disagreement with Eq.~\ref{assumptionB}, which authorizes detection in $B$ or $C$. So there's a contradiction and we must conclude that it's impossible to have such a configuration and  therefore Eq.~\ref{assumptionB} can not be true in situations involving  several points on the same trajectory!\\
\indent This clearly undermines DD's position \cite{DasDurr2018,DasDurr2019} that the presence of a detector should not be taken into account when analyzing arrival times (it also undermines some predictions made in \cite{Rafsanjani2023} concerning backflow). In fact, we are faced with two alternatives:\\ 
\indent 1) Either we use detectors operating in the weak coupling regime, but then we have to amend the implicit assumption that any measurement is based solely on a POVM, and we have to add a post-selection and filtering condition (post analysis) taking into account the dBB dynamics. Also in this regime (at least if we neglect spin) $\mathcal{P}_{\textrm{detec.}}^\Psi(\Sigma,\tau)\simeq  \Sigma d N  \sigma_{ext}|\mathbf{J}^{\Psi^{(0)}}|\neq \mathcal{P}_{\textrm{dBB}}^{\Psi^{(0)}}(\Sigma,\tau)$. \\
\indent 2) Or we use a strongly coupled detection regime, but then we generally have to abandon the idea of being able to directly measure the dBB distribution of arrival times based on Eq. \ref{dBBflux}.
In case 2 we could of course eliminate  the problem by limiting the analysis to the detection at only one single space-time point located in the small region $\delta  \Omega$. But in turn this  would  mean that a single experiment could not measure all the distribution $\mathcal{P}_{\textrm{dBB}}^{\Psi^{(0)}}(\Sigma,\tau)$. We would need several experiments in order to reconstruct the distribution of arrival times.  Additionally, in this situation nothing would prohibit us to record the distribution, at say, point B of Figure \ref{Fig4} (b). Indeed,  since there is no detector at A there is no scattering from region A disturbing the local motion at B.  Therefore like in the weak-coupling regime  we need to add a post-selection  depending on the full  dBB dynamics in order to filter out such detection events. Like for the weak coupling regime this clearly contradicts the assumption of an only-POVM-based quantum measurement procedure.\\
\indent There are other issues that we have mentioned and which we must now consider as they play a central role in the analysis of the work of DD and GTZ. Indeed, DD's predictions involve spin $1/2$ particles, so we need to include a magnetic current term (see Eq.~\ref{currentspin}) in our analysis. As we shall see, this has a strong impact on both regime 1 (weak coupling) and regime 2 (strong coupling).
\subsection{The spin-dependent problem and the measurement of the  first arrival time distribution of Das and D\"{u}rr}  
\indent In the previous analysis we didn't include the spin-$1/2$ required in the work of DD \cite{DasDurr2018,DasDurr2019} and GTZ \cite{GTZ,GTZ2}.  For this purpose we need to consider the dynamics of electron using either the Pauli or Dirac wave equation.  Taking the relativistic Dirac wave equation we have for the electron bi-spinor $\Psi(x)\in \mathbb{C}^4$: 
\begin{eqnarray}
i\gamma^\mu\partial_\mu \Psi(x)=m\Psi(x)+ e\gamma^\mu A_\mu(x) \Psi(x)\label{Dirac}
\end{eqnarray} where $\gamma^0=\beta$ and $\boldsymbol{\gamma}=\beta\boldsymbol{\alpha}$ are Dirac matrices, $e=-|e|$ the electron charge and $A_\mu(x)$ the external electromagnetic field at space-time point $x:=[t,\mathbf{x}]$. By an obvious generalization of the previeous non-relativistic analysis we can define absorbing detector involving complex 4-vector potential $A^\mu:=[A^0=\Phi,\mathbf{A}]=\textrm{Re}[A^\mu_{eff}]+i\textrm{Im}[A^\mu_{eff}]$. The local conservation law for the 4-current $J^\mu=\overline{\Psi}\gamma^\mu\Psi:=[J^0=\rho^\Psi=\Psi^\dagger\Psi,\mathbf{J}^\Psi=\Psi^\dagger\boldsymbol{\alpha}\Psi]$ is deduced from Eq.~\ref{Dirac} and reads
\begin{eqnarray}
-\partial_\mu J^\mu:=-\partial_t\rho^\Psi-\boldsymbol{\nabla}\cdot\mathbf{J}^\Psi=-2e\textrm{Im}[A^\mu_{eff}]J_\mu=-2e\textrm{Im}[\Phi_{eff}]\rho^\Psi+2e\textrm{Im}[\mathbf{A}_{eff}]\cdot\mathbf{J}^\Psi\label{Diraccur}
\end{eqnarray} 
which generalizes Eq.~\ref{fluxdis} obtained in the non-relativistic regime  for spinless particles.\\
\indent From this relation we can (apriori) extend the analysis of Section \ref{section6} and define the probability of absorption by a detector located in the volume $\delta \Omega$   
\begin{eqnarray}
\mathcal{P}_{\textrm{detec.}}^\Psi(\delta \Omega)=-2e\int_{\delta\Omega}d^4x\textrm{Im}[A^\mu_{eff}]J_\mu(\mathbf{x},t) \label{detex4Dre}
\end{eqnarray} which generalizes Eq.~\ref{detex4D} obtained in the non-relativistic regime for spinless particles.
 However, unlike in the non-relativistic regime, there is an obvious difficulty: the scalar product $-e\textrm{Im}[A^\mu_{eff}]J_\mu(\mathbf{x},t)$ has no imposed sign. More precisely, in the non-relativistic regime, the quantity $-\textrm{Im}[V_{eff}]|\Psi|^2(\mathbf{x},t)$ could always be positive if the medium obeys a natural causal and entropic condition $-\textrm{Im}[V_{eff}]\geq 0$ associated with inelastic scattering and dissipation in the medium (i.e., due to coupling with a thermal bath).  Of course, media with gain (producing particles) such that $-\textrm{Im}[V_{eff}]\leq 0$ were also potentially possible, but we could always imagine making a choice between the two alternatives. In the case of lossy media, interpreting Eq.~\ref{detex4D} as the probability of absorption associated  with a POVM was therefore straightforward. In the relativistic regime we can not in general be sure that the intrinsic  properties of the medium will always impose a value of $-e\textrm{Im}[A^\mu_{eff}]J_\mu(\mathbf{x},t)$ strictly positive or negative. It means that for a given field $\textrm{Im}[A^\mu_{eff}]$ $\mathcal{P}_{\textrm{detec.}}^\Psi(\delta \Omega)\geq 0$ (loss) or $\mathcal{P}_{\textrm{detec.}}^\Psi(\delta \Omega)\leq 0$ (gain) depending on the wave function $\Psi$.  Therefore, Eq.~\ref{detex4Dre} is not generally a POVM. Still this quantity has always a physical meaning: If  $\mathcal{P}_{\textrm{detec.}}^\Psi(\delta \Omega)\geq 0$ it represents a probability of absorption and  alternatively  if $\mathcal{P}_{\textrm{detec.}}^\Psi(\delta \Omega)\leq 0$ then $-\mathcal{P}_{\textrm{detec.}}^\Psi(\delta \Omega)\geq 0$  represents a probability of emission (gain). The crux is that it depends explicitely on the wave function $\Psi$ used.   This implies that in the relativistic regime the concept of POVM must be used cautiously. We stress that the present analysis is in line with works applying dBB theoy to quantum field theory (QFT) where source/sink terms must be added in order to explain creation/annihilation of particles by fields. \cite{QFT} \\       
\indent Furthermore, supposing that for a given wave function $\Psi$ we have indeed $\mathcal{P}_{\textrm{detec.}}^\Psi(\delta \Omega)\geq 0$ and that we are considering a strongly efficient detector we must have as in Section \ref{section6}: \begin{eqnarray}
\mathcal{P}_{\textrm{detec.}}^\Psi(\delta \Omega)=-2e\int_{\delta\Omega}d^4x\textrm{Im}[A^\mu_{eff}]J_\mu(\mathbf{x},t)
\simeq \delta t\int_{\Sigma_{in}} d^2\Sigma_\mathbf{x}\mathbf{n}_\mathbf{x} \cdot\mathbf{J}^\Psi(\mathbf{x},t)
\label{detex4Dbisre} 
\end{eqnarray} where the surviving contribution comes from a surface integral on the entrance side of the detector (there is no 	ambiguity here since the absorption condition $\mathcal{P}_{\textrm{detec.}}^\Psi(\delta \Omega)\geq 0$ fixes the direction of the decay of the wavec function inside the detector).\\  
\indent However, there is now a new difficulty. Indeed, the Dirac current $J^\mu$ can be separated into a convective current, a magnetic term  and an absorbing term using the so called Gordon formula:
\begin{eqnarray}
J_\mu=\frac{i}{2m}(\overline{\Psi}\stackrel{\textstyle\leftrightarrow}{\rm D}_{\mu}\Psi)- \frac{1}{2m}\partial^\nu(\overline{\Psi}\sigma_{\nu\mu}\Psi)-\frac{e}{m}\overline{\Psi}\sigma_{\mu\nu}\Psi\textrm{Im}[A^\nu_{eff}]\label{dirac1}
\end{eqnarray}  with $\sigma_{\nu\mu}=\frac{i}{2}[\gamma_\nu,\gamma_\mu]$, $D^\mu=\partial^\mu+ie\textrm{Re}[A^\mu_{eff}]$. The absorbing  term  is usually not present since the electromagnetic field is supposed to be real valued.  Here this is not the case and we must in general include this contribution.  In the non relativistic limit this reduces to $\rho^\Psi\simeq \Psi^\dagger\Psi$  and 
\begin{eqnarray}
\mathbf{J}^\Psi\simeq \frac{1}{2m}(\Psi^\dagger\stackrel{\textstyle\leftrightarrow}{\rm \boldsymbol{\pi}}\Psi)+\frac{1}{2m}\boldsymbol{\nabla}\times[\Psi^\dagger \boldsymbol{\sigma} \Psi] +\frac{e}{m}\textrm{Im}[\mathbf{A}_{eff}]\times \Psi^\dagger \boldsymbol{\sigma} \Psi  \label{dirac2}
\end{eqnarray}
where  $\boldsymbol{\pi}=\frac{\boldsymbol{\nabla}}{i}-e\textrm{Re}[\mathbf{A}_{eff}]$ and $\Psi\in \mathbb{C}^2$ is now a bispinor. Now the problem is that the divergence of the magnetic term in Eq.~\ref{dirac1} or \ref{dirac2} cancels out and consequently the associated surface integral calculated  using Gauss's theorem  over a closed surface surrounding the detector region (in the 3D or 4D formalism) vanishes in Eq.~\ref{detex4Dbisre}. Therefore, in the non-relativistic regime Eq.~\ref{detex4Dbisre} actually reduces to    
\begin{eqnarray}
\mathcal{P}_{\textrm{detec.}}^\Psi(\delta \Omega)\simeq \delta t\int_{\Sigma_{in}} d^2\Sigma_\mathbf{x}\mathbf{n}_\mathbf{x} \cdot[\frac{1}{2m}(\Psi^\dagger\stackrel{\textstyle\leftrightarrow}{\rm \boldsymbol{\pi}}\Psi)+\frac{e}{m}\textrm{Im}[\mathbf{A}_{eff}]\times \Psi^\dagger \boldsymbol{\sigma} \Psi ].
\label{detex4Dbisre} 
\end{eqnarray}
The most important consequence is that an efficient detector (i.e., $\eta \sim 1$) cannot register a signal proportional to the total current flow: the detector is not sensitive to the magnetic term $\frac{1}{2m}\boldsymbol{\nabla}\times[\Psi^\dagger \boldsymbol{\sigma} \Psi]$.  But it is precisely this term that plays a crucial role in the analysis of DD and GTZ, with disastrous consequences for the analysis of DD in this regime.  More precisely, in the case of a simple detector where the coupling is via the scalar field $V_{eff}:=e\Phi_{eff}$ ($A_{eff}^\mu:=[\Phi_{eff},0]$),  we recover the analysis made in the previous sections for spinless particles (indeed we  can always impose $-2e\textrm{Im}[\Phi_{eff}]\rho^\Psi\geq 0$ imposing $\mathcal{P}_{\textrm{detec.}}^\Psi(\delta \Omega)\geq 0$), but now we see that the detector will only be sensitive to the convective term which in the example of DD reads
\begin{eqnarray}
\mathbf{J}_{conv.}^{\Psi_{\hat{\mathbf{s}}}}(\mathbf{x},t)=|\Phi(\rho,z,t)|^2\frac{\boldsymbol{\nabla}S(\rho,z,t)}{m}
\end{eqnarray} without the magnetic term $\frac{-1}{m}\hat{\mathbf{s}}\times\boldsymbol{\nabla}|\Phi(\rho,z,t)|^2$ of Eq.~\ref{currentspin}. In the work of DD \cite{DasDurr2019} an explicit formula is given for the current which in their system of normalized units (see Eq. 18 in \cite{DasDurr2019}) reads:
\begin{eqnarray}
\mathbf{J}_{conv.}^{\Psi_{\hat{\mathbf{s}}}}(\mathbf{x},t)=|\Phi(\rho,z,t)|^2\frac{tz}{1+tz^2}\hat{\mathbf{z}}\label{convec}
\end{eqnarray}  which is directed parallel to the axis of the waveguide. It is interesting to note that a dBB dynamics without any magnetic term  is often connsidered as a good alternative (see discussions in \cite{Holland,BohmHiley}). We can debate endlessly the motivations for the different dynamics but in the end we see that the detectors are in the present regime ignoring the spin term.  The  dBB trajectories deduced from this truncated current are thus straight lines parallel to the $z$ axis  and the velocity of the particle is $\frac{tz}{1+tz^2}\hat{\mathbf{z}}$. In this regime  there is never back-flow. Of course for a Bohmian the question of which velocity is the good one is fundamental but from the point of view of detection theory   only the convective term plays a role and the trunctated dynamics without spin term is the only relevant one.\\   
\indent Most importantly,  this convective term generates a probability distribution of first arrival times that is independent of the orientation of the incident  spin $\hat{\mathbf{s}}$. The analytical formula is again given in \cite{DasDurr2019} for the convective current of Eq.~\ref{convec} (see Eqs.~51-52 in \cite{DasDurr2019}) and we have:
 \begin{eqnarray}
\mathcal{P}_{\textrm{detec.}}^\Psi(\tau,L)= \frac{4L^3}{\lambda_0 \sqrt{\pi}}\frac{\tau e^{-\frac{L^2}{1+\tau^2}}}{(1+\tau^2)^{5/2}}
\label{detexDD} 
\end{eqnarray} where $z=L$ is the position of the detector localized along a cross-section of the wave guide and $\lambda_0$ is a constant \cite{DasDurr2019}.   \\
\indent We stress that for DD the distribution of first arrival times given by Eq.~\ref{detexDD} is only valid in the longitudinal cases  where $\hat{\mathbf{s}}=\pm \hat{\mathbf{z}}$ whereas here in presence of strongly efficient detectors this distribution is actually valid for every spin orientation $\hat{\mathbf{s}}$! In particular this distribution is never vanishing, i.e., there is no critical time $\tau_{max}$ for which $\mathcal{P}_{\textrm{detec.}}^\Psi(\tau,L)=0$ for $\tau> \tau_{max}$. This is very different from the predictions obtained by DD \cite{DasDurr2018,DasDurr2019} with transverse spins. Clearly this implies that assuming the strong coupling regime all paradoxes of DD and GTZ disappear.\\
\indent More precisely, in the strong coupling regime, where detectors are inherently sensitive only to the first arrival time, we see that GTZ's analysis  \cite{GTZ} is clearly validated to the detriment of DD's conclusions \cite{DasDurr2018}. Indeed, in this regime,  Eq. ~\ref{sum}  is trivially true, as there is no longer any spin dependence ($\mathcal{P}_{\textrm{detec.}}^\Psi(\tau,L)$ is for all practical purposes a POVM). Bell's theorem is also safe: It is not possible to use this type of experiment to send faster-than-light signals, as the distribution is invariant to spin basis shifting. Again this agrees with GTZ \cite{GTZ}.
   
At this stage, you could say the die was cast: GTZ were right and DD were wrong.  However, we shouldn't jump to conclusions. We haven't yet analyzed the problem in terms of weak coupling detection.    

As we said, this weak coupling regime has two inherent shortcomings: the first concerns the notion of POVM, which the approach seems to cast doubt on, since the notion of dBB trajectory must be taken into account in order to make arrival time predictions. The second problem stems from the fact that the $\mathcal{P}_{\textrm{dBB}}^{\Psi^{(0)}}(\Sigma,\tau)$ distribution measured in the non-relativistic regime depends only on the norm of the probability current $\mathcal{P}_{\textrm{detec. }}^\Psi(\Sigma,\tau)\simeq \Sigma d N \sigma_{ext}|\mathbf{J}^{\Psi^{(0)}}|$. However, the second problem was obtained in the context of a non-relativistic theory for spinless particles.  In the context of Dirac or Pauli theory, this problem can in fact be corrected. Indeed, starting from Eq.~\ref{detex4Dre} we deduce that in general we have
\begin{eqnarray}
\mathcal{P}_{\textrm{detec.}}^\Psi(\delta \Omega)=-2e\int_{\delta\Omega}d^4x\left(\textrm{Im}[\Phi_{eff}]\Psi^\dagger\Psi(\mathbf{x},t)-\textrm{Im}[\mathbf{A}_{eff}]\cdot\Psi^\dagger\boldsymbol{\alpha}\Psi(\mathbf{x},t)\right) \label{detex4Dreweak}
\end{eqnarray}  where $\Psi$ is a Dirac bispinor. In the spinless case only the first term appears and in the weak coupling regime we have indeed an absorption probability  proportional to $\Psi^\dagger\Psi(\mathbf{x},t)$, i.e. to the density of probability. In the Dirac-Pauli theory this is possible if the detector is scalar, i.e., if $\textrm{Im}[\mathbf{A}_{eff}]=0$. But for a spin$-1/2$ particle this is not the only option. We can in principle develop experimental configuration with absorbing field such that $\textrm{Im}[\Phi_{eff}]=0$ but $\textrm{Im}[\mathbf{A}_{eff}]\neq 0$. In this alternative the `probability' of detection/gain reads     
\begin{eqnarray}
\mathcal{P}_{\textrm{detec.}}^\Psi(\delta \Omega)=+2e\int_{\delta\Omega}d^4x\textrm{Im}[\mathbf{A}_{eff}]\cdot\Psi^\dagger\boldsymbol{\alpha}\Psi(\mathbf{x},t)\sim \epsilon \delta t\delta V\mathbf{n}\cdot\mathbf{J}_{total}^{\Psi}(\mathbf{x}_{\textrm{detec.}},t_{\textrm{detec.}}) \label{detex4DreweakBis}
\end{eqnarray} with $\epsilon \mathbf{n}\sim +\frac{2}{\delta V} \int_{\delta V}d^3x\textrm{Im}[\mathbf{A}_{eff}(\mathbf{x},t)] $ a coupling efficiency (compare with Eq.~\ref{badone}). This `probability' of local absorption/gain depends on the full Dirac current at the detector position and thus in principle it is possible to build such a detector which in the weak-coupling regime would give a signal directly related  to the dBB probability predicted by DD \cite{DasDurr2018,DasDurr2019}. Moreover, in order to have absorption and not gain it means that we must define locally the field $\textrm{Im}[\mathbf{A}_{eff}(\mathbf{x},t)]$ in order to have  $\mathcal{P}_{\textrm{detec.}}^\Psi(\delta \Omega)\geq 0$ in the small 4-volume $\delta \Omega$. This procedure is wave function dependent since for a given field $\textrm{Im}[\mathbf{A}_{eff}(\mathbf{x},t)]$ we cannot impose the sign of Eq.~\ref{detex4DreweakBis} for every wave functions! Again $\mathcal{P}_{\textrm{detec.}}^\Psi(\delta \Omega)$ given  by Eq.~\ref{detex4Dreweak} is not generally a POVM but still its physical interpretation  in term or loss or gain is obvious. The fact that it strongly depends on $\Psi$ show once more that in the relativistic regime local interactions don't  simply lead to POVM. Yet, by specifically and locally engineering  a field $\textrm{Im}[\mathbf{A}_{eff}(\mathbf{x},t)]$ we can imagine to develop detectors adapted to a given  wave function $\Psi$.\\
\indent Note, however, that at the very end of a measurement process, POVMs are still used.  A localized detector with $\mathcal{P}_{\textrm{detec.}}^\Psi(\delta \Omega)>0$ or $\mathcal{P}_{\textrm{detec.}}^\Psi(\delta \Omega)<$ properties will behave either as an absorber or as an emitter, depending on the case and the $\Psi$ wave function chosen.   If we know apriori by calculation (i.e. dBB trajectories) how it will behave, then in the end the experimenter will either have to count absorbed particles (e.g., trapped in the detectors or channelled to particle-counting outputs in the far-field), or emitted particles once again sent and redirected to more conventional counters in the far-field regime.   In these final counting regimes, we ultimately end up using $\int_\Delta dq |q\rangle\langle q|$ projectors associated with POVMs in given regions of space $\Delta$ (i.e., in the far-field). The fact that dBB theory is ultimately based on such spatial location and counting experiments was already pointed out by de Broglie and Bohm and justifies the use of POVMs.  Clearly, however, there is no universal detection procedure, and we need to add elements foreign to POVMs in order to do good dBB physics. \\
\indent Of course, we're just here building a proof of principle, and we can see that there's nothing in the laws of physics to prevent the construction of such detectors. However, clearly more work needs to be done to define a precise and efficient design for such a Dirac or Pauli current detector that would allow us to trace the Bohmian distribution of arrival times predicted by DD. This goes far beyond the scope of the present study. Now we saw that POVM are not the end ot the story but before concluding, however, it remains to return to the second objection against the weak coupling regime: namely, the weakening of the exclusive use of POVMs, which is assumed as a postulate by GTZ among others and is associated with  a post-analysis of the data. 

\subsection{Beyond the standard only-POVM-based quantum measurement procedure }
\indent The fact that the notion of POVM appears in any probabilistic analysis in quantum mechanics and more specifically in Bohmian mechanics is not surprising, as we pointed out at the beginning of this article. However, the belief that one can limit oneself to using POVMs to interpret quantum experiments within the framework of dBB theory is based on prejudiced beliefs and demonstrates that a better understanding of the importance of the Bohmian approach can be obtained. The problem  is actually much more general that the one we analyzed in the previous subsection and which concerned only the Dirac-Pauli equation and the relativistic regime (or the regime of the Pauli equation with spin).\\
\indent Let's return to the example of the two-slit experiment discussed at the beginning of this article in Section \ref{section2}. As we showed then, it is possible, thanks to the dBB theory, to retrodict the passage of the particle through one or other of the apertures while detecting interference fringes.   To do this, we need to know precisely the shape of the wave function $\Psi(\mathbf{x},t)$ used in the experiment (and in particular its phase) in order to calculate the Bohmian velocity field and thus obtain the ‘which path’ information.   Of course, the whole method relies on the validity of Born's rule on quantum probabilities, and in the end this necessarily implies the use of POVM in the analysis. However, knowledge of the theoretical Bohmian trajectory allows us to find information that we would say is ‘hidden’ if we didn't know this dBB theory. In other words, we start with the raw measurements using Born's rule, but we have to carry out post-analysis or filtering to process the data and highlight the correct Bohmian information.  This new methodology has been strongly advocated in recent years by Detlef D\"{u}rr and I consider this point to be a major contribution to our understanding of Bohmian theory. Clearly this is far from being accepted by the whole ‘Bohmian’ community but the lack of consensus shows in my opinion even more the importance of the classical physics prejudices that have survived among the dBB community.\\
\indent D\"{u}rr often illustrated his argument by using Einstein's reply quoted in Section \ref{section2}:
\begin{quote}
`\emph{it is the theory which alone decides what is measurable}'.\cite{Heisenberg}
\end{quote}
 which Einstein gave to Heisenberg in 1926 when the latter  claimed that he could build a theory using only the notion of the observable.   What Einstein reminded or taught Heisenberg \cite{Heisenberg} was that every scientific theory begins with a quasi-metaphysical act: a theory has to be postulated and this act, although motivated by previous observations, is free.   Then comes the prediction, and empirical data can only be interpreted within a precise theoretical framework.  This is the heart of the hypothetico-deductive method advocated by Boltzmann and Einstein. Here, we are interested in the dBB theory and therefore following Einstein's hypothetico-deductive method the analysis of data must include the Bohmian dynamics in order to be predictive.\\
 \indent This is clearly the case for the retrodiction obtained in the two-slit experiment, which enabled us to trace back to the which path information thanks to precise knowledge of the dBB trajectorie (interestingly, it was only by forgetting the fundamentally quantum character of these dBB trajectories that Heisenberg and many others after him thoughtlessly deduced that Bohmian dynamics was surreal and that not trajectory interpretation could explain wave-particle duality). This is also clearly the case here (and in agreement with the conclusions of D\"{u}rr and Das \cite{DasDurr2018,DasDurr2019}) for the analysis of the first arrival times of particles on a detector. Going back to the above analysis of strong and weak regimes, what we have deduced is indeed the need to explicitly take into account the Bohmian dynamics in order to be predictive and to reconstruct the first arrival time probability $\mathcal{P}_{dBB}^\Psi(\mathbf{x},t)$ from the raw data.\\
\indent This suggests the following experimental scenario for measuring the arrival time distribution predicted by DD:
Use a set of detectors operating in the weak coupling regime in agreement with Eq.~\ref{detex4DreweakBis} and sensitive to the probability current including the spin term. This set of detectors is distributed in space-time in such a way as to map the probability density of Bohmian arrival times $\mathcal{P}_{\textrm{detec}}^{\Psi^{(0)}}(\Sigma,\tau)\geq 0$. This requires a specific engineering of the local fields $\textrm{Im}[\mathbf{A}_{eff}(x_i)]$ in each spatio-temporal regions $\delta \Omega_i$ where are located the detectors in order to impose $\mathcal{P}_{\textrm{detec}}^{\Psi^{(0)}}(\Sigma,\tau)\geq 0$.  As we are working in the weak coupling regime, the initial wavefunction is very weakly perturbed, enabling us to carry out a single experiment without changing the protocol from point to point in space-time.   However, in this weak coupling regime we also need to perform a post-analysis to filter out signals that may or may not be associated with first arrival times, second arrival times etc... This is clearly wave function dependent and shows that elements foreign to POVM must be considered.  Although the method is based on good absorbing detectors, it can only be interpreted physically if this post-analysis is carried out.   This is very much in line with the principle described above in the two-slit example. \\
\indent It could be argued that if knowledge of Bohmian trajectories is necessary in order to carry out this post-analysis or post-selection this would be a purely theoretical element based on unobservable trajectories, i.e. `hidden variables'.  In reality, dBB trajectories are not unobservable in principle. Weak measurement protocols \cite{WM1} can be used to map the velocity field  \cite{WM2,WM3,WM4} and then trace the trajectories followed by the particles. In principle, then, we could imagine a pre-experiment that would first map dBB trajectories for the $\Psi^{(0)}_t$ initial wave function we are interested in (in principle this could also involve the regions of detectors but this would  be very difficult since it would require near-field measurements). Only once we know these dBB trajectories can we carry out the post-selection required in our arrival time experiment to reconstruct the Bohmian probability distribution.\\
\indent Clearly, with this protocol we can bypass the objections raised by GTZ \cite{GTZ,GTZ2}. First of all, Eq. \ref{sum} presupposed that arrival time measurements were entirely based on the notion of POVM. However, although our detectors here are fundamentally absorbing since we have $\mathcal{P}_{\textrm{detec}}^{\Psi^{(0)}}(\Sigma,\tau)\geq 0$, detector engineering  together with post-selection are specifically Bohmian and non-linear (they vary strongly with the wave functions used).  Remarkably, the present protocol while not based only on POVM is physically acceptable and doesn't contradict any fundamental law.\\ 
\indent Secondly, the objections related to Bell's theorem and no-signalling are also answered. Indeed, in the 'Bell-Maudlin-Das experiment' described in section 2, it is assumed that Bob can measure the distribution of dBB arrival times independently of knowledge of the spin of his particles. This makes sense in a procedure based solely on POVMs. But here the design of the detectors and the method of post-selecting the data require knowledge of the wave function and therefore of the particle spin.  Bob can't establish the $\mathcal{P}_{\textrm{dBB}}^{\Psi^{(0)}}(\Sigma,\tau)$ probability distribution without prior knowledge of the wave function and spin of the objects measured by Alice. Moreover, Bob could still decide to use a fixed setup such that the potentials $\textrm{Im}[\mathbf{A}_{eff}(x_i)]$ at the various points of the detector is uniquely defined for all the wave functions. If he is fixing the set up in such a way then the various `probabilities'   $\mathcal{P}_{\textrm{detec.}}^\Psi(\delta \Omega)$ for each elementary volumes $\delta \Omega$ of the detector are not necessarily positive (this is reminiscent of the presence or loss and gain in the general dynamics). This is the case in particular if we have backflow as in DD setup \cite{DasDurr2018}.  The detector is thus not always working correctly and sometimes part of the full detector emits particles instead of absorbing them. Moreover, if we consider the full Dirac current with a convective and magnetic contributions in the configuration developed by DD \cite{DasDurr2019} (see Eq.~\ref{currentspin}) we can easily prove (see Appendix \ref{appendix5}) that the full integrated signal $\mathcal{P}_{\textrm{detec.,full}}^\Psi(\Sigma,t)\simeq \eta\int d^2\Sigma_{\mathbf{x}}J_z^{\Psi_{\hat{\mathbf{s}}}}(\mathbf{x},t)$ recorded by the detector without post-selection         
is given by 
 \begin{eqnarray}
\mathcal{P}_{\textrm{detec.,full signal }}^\Psi(\Sigma,\tau)
=\eta \frac{4L^3}{\lambda_0 \sqrt{\pi}}\frac{\tau e^{-\frac{L^2}{1+\tau^2}}}{(1+\tau^2)^{5/2}}.
\end{eqnarray} with $\eta\ll 1$
 This is precisely (up to the $\eta$ coefficient) the spin independent distribution considered in  \cite{DasDurr2019} (see Eq.~\ref{detexDD}). In other words: the signal being spin independent it can not be used to violate non signalling and sent a signal. Bob can of course decide to post-select data in order to reconstruct the first arrival dBB distribution (which is spin dependent). But in order to do that he must already know  what is the spin measured by Alice in order to correlate the information!\\
 \indent To sum up and conclude, in this work we have analyzed in detail DD's proposal \cite{DasDurr2018,DasDurr2019} to measure particle arrival times using dBB theory. We have compared their work with the criticism made by GTZ. \cite{GTZ,GTZ2} To this end, we have studied in detail the notion of particle detection in quantum mechanics in the context of DBB theory. We concluded that both DD \cite{DasDurr2018,DasDurr2019} and GTZ \cite{GTZ,GTZ2} were both right and wrong. More specifically, DD were right in believing that their specific Bohmian predictions involving the back-flow phenomenon could be observed. However, they were wrong to believe that the impact of detector physics could be neglected in their analyses. To be sure, the dBB $\mathcal{P}_{\textrm{dBB}}^{\Psi^{(0)}}(\Sigma,\tau)$  distribution is merely an ideal, theoretical formulation of particle flow in space-time.  However, only the $\mathcal{P}_{\textrm{detec.}}^\Psi(\delta \Omega)$ probability associated with absorption or, more generally, interaction phenomena makes sense in the context of a complete physical theory, and dBB theory as such is no exception to this fundamental fact.\\
\indent In this work, we have clearly demonstrated the existence of two regimes: weak and strong coupling, corresponding to low and high detection or absorption efficiency respectively. The strong coupling regime is the most natural, as it corresponds to the experimenter's natural expectation and it will lead to first arrival time distributions. In this regime the  detection probability $\mathcal{P}_{\textrm{detec.}}^\Psi(\delta \Omega)$ (wich is a POVM in the non-relativistic regime) reduces approximately to the  dBB probability Eq.~\ref{dBBflux} $\mathcal{P}_{dBB}^\Psi(\mathbf{x},t) :=|\mathbf{J}^\Psi(\mathbf{x},t)\cdot\mathbf{n}(\mathbf{x})|d\Sigma_\mathbf{x}$ which  is not a POVM. Since this is true only for some wave functions $\Psi$ there is no paradox. However, the method  is highly invasive and strongly disturbs the wave function and dBB trajectories, which in general can lead to major technical difficulties. Therefore, it would be impossible to engineer complex time arrival detectors adapted to several wave functions $\Psi_1,\Psi_2,...$  some presenting back-flow some others not.   What's more, in the relativistic domain (requiring the Dirac equation) or in the Pauli equation regime for spin 1/2 electrons, the notion of POVM is even more difficult to apply and we have seen that it's very hard to make a measurement approaching the dBB prediction because the spin magnetic current is generally undetected.\\   
\indent We deduced that the weak coupling regime was ultimately more appropriate for measuring the probability distribution predicted by DD. However, there is a price to pay. First of all, we must learn to give up the tenacious belief that only physics based on the notion of POVM has the right to quote. In fact, in the dBB framework, it is necessary to abandon this prejudice as soon as we seek to analyze trajectories (as we have shown with several examples).   In keeping with Einstein's credo that `only theory decides what is to be measured', we have shown that, in order to measure the $\mathcal{P}_{\textrm{dBB}}^{\Psi_{\hat{\mathbf{s}}}}(\Sigma,\tau)$ probability distribution predicted by DD \cite{DasDurr2018,DasDurr2019} (and which depends on spin orientation $\hat{\mathbf{s}}$), in the weak coupling regime we must necessarily carry out a post-analysis or post-selection to filter and classify the events detected, corresponding to first detection, second detection, etc. This point is fundamental and strongly contradicts GTZ's conclusions, which rely solely on the notion of POVM in their critical analysis.\\
\indent In the end, however, we agree with GTZ \cite{GTZ,GTZ2} on two points: firstly, the physics of the detector cannot be neglected in the analysis, as pointed out above (although this actually constitutes a weaker agreement with GTZ, who hastily concluded that no dBB arrival time measurement was possible based on their POVM analysis, whereas we demonstrate the opposite here); secondly, it is impossible within the framework of `standard' non modified  quantum mechanics, in which the dBB theory is embedded (i.e. without questioning the foundations and without adding new physics at a `sub-quantum' level), to contradict the results of Bell's theorem and violate the no-signalling condition. In fact, an analysis obtained within the framework of the weak coupling regime shows that DD and Maudlin's proposal could not lead to such violations and to a hypothetical transmission of detectable supraluminal information. In our view, this is fortunate, as it means that the dBB theory is still completely empirically equivalent to the orthodox approach (in areas where these approaches are comparable). Of course, new physics is always possible, \cite{DasAris} but it is by no means necessary here to agree with the results of DD and GTZ.  \cite{DasDurr2018,DasDurr2019,GTZ,GTZ2} For example, as indicated in the introduction, it is in principle possible within the framework of dBB theory to relax Born's rule, i.e. to abandon quantum equilibrium. In this regime, it would be possible, in principle, to transmit superluminal signals to build a `Bell telephone'. But this remains highly speculative and could have a line with primordial cosmology as proposed for instance by Valentini. \cite{nonsigValentini,nonsigValentini2} \\
\indent At a more fundamental level, our work should not be seen as an attempt to prove the correctness or truth of the dBB interpretation (contrary to hypotheses that have been discussed in the past \cite{Cushing1995}). As we have shown, the dBB theory fits in very well with the theoretical framework of quantum mechanics, and allows us to recover all its empirical content. In this field of measurement theory, Bohmian and orthodox quantum mechanics are empirically equivalent.  Of course, the ontological clarity and absence of a measurement problem (i.e. the absence of a wavefunction collapse) is a great advantage for dBB theory. However, other ontological approaches could undoubtedly predict other trajectories and at the same time account for arrival time experiments. Also, as mentionned in the introduction we could just add an arbitrary  $\boldsymbol{\nabla}\times \mathbf{F}(\mathbf{x},t)$ term to the local current in order to obtain a new Bohmian ontology. The general methodology here would be to develop detectors adapted to these new probability currents and dBB dynamics. This would clearly define new distribution of probability for the arrival times and we see no reason or physical law which could prohibit to imagine detectors for such alternative theories.   From a philosophical point of view, this leads us to be more modest about our preferred theories, while at the same time encouraging more comparative analysis of different approaches. \cite{Muga2000}\\

\indent I acknowledge useful discusssions with Tim Maudlin and Jean Bricmon concerning the role of Born's rule in the derivation of the no-signalling theorem.

\appendix
\section[\appendixname~\thesection]{POVM and dBB theory}
\label{appendix0}
\indent In order to describe a quantum measurement we start with a subsystem $S$ wave function $|\psi_0^{S}\rangle\in \mathcal{H}^S=\sum_n c_n|n^{S}\rangle$ expanded in a complete basis $|n^{S}\rangle$ and initially uncoupled to a pointer $M$ wave function $|\Phi_0^{M}\rangle\in \mathcal{H}^M$.  During a generalized von-Neumann measurement the interaction between $S$ and $M$ is characterized by an unitary evolution operator $\hat{U}^{SM}$  acting on the full the Hilbert space $\mathcal{H}^S\otimes\mathcal{H}^M$ and it leads to entanglement:
\begin{eqnarray}
|\Psi_0^{SM}\rangle=|\psi_0^{S}\rangle|\Phi_0^{M}\rangle=(\sum_n c_n|n^{S}\rangle)|\Phi_0^{M}\rangle
\xrightarrow{\hat{U}^{SM}} |\Psi_t^{SM}\rangle=\sum_n c_n|\Psi_n^{SM}\rangle\label{POVM1}
\end{eqnarray} where we have $\langle n^S|m^S\rangle=\delta_{nm} \rightarrow\langle \Psi_n^{SM}|\Psi_m^{SM}\rangle=\delta_{nm}$.  We stress that in standard projective von-Neumann measurements  $|\Psi_t^{SM}\rangle=|n^{S}\rangle|\Phi_t^{M}\rangle$ but here we consider a more general case.
In the dBB framework the physical probabilities are defined in the configuration space and therefore if $q$ are the spatial coordinates for the $S$ sub-system and $\xi$ the spatial coordinates for the  $M$ sub-system  we have initially the wave function $\Psi_0^{SM} (q,\xi)$ which evolves  as $\Psi^{SM} (q,\xi,t)$ at time $t$. We can thus rewrite 
\begin{eqnarray}
\Psi_0^{SM}(q,\xi)=\psi_0^{S}(q)\Phi_0^{M}(\xi)=(\sum_n c_n \psi_n^{S}(q)\Phi_0^{M}(\xi)
\xrightarrow{\hat{U}^{SM}} \Psi^{SM}(q,\xi)=\sum_n c_n\Psi_n^{SM}(q,\xi,t)
\end{eqnarray}  
In order for an observer to legitimately speak of a quantum measurement, the physical variables of the pointer, which in the dBB framework are necessarily  the coordinates $\xi$,  $M$ must move by an observable quantity in such a way that we can experimentally distinguish the different eigenvalues $n$, $m$,... associated with the different states $|n^{S}\rangle$, $|m^{S}\rangle$,... To do that in a non-ambiguous way  we must be sure that the different waves functions  $\Psi_n^{SM}(q,\xi,t)$, $\Psi_m^{SM}(q,\xi,t)$,... are non-overlapping in the $\xi-$configuration space.   In other words, these wave functions $\Psi_n^{SM}(q,\xi,t)$, $\Psi_m^{SM}(q,\xi,t)$,... must have finite disjoint supports $\Delta_n$, $\Delta_m$,... in the $\xi-$configuration space such that
\begin{eqnarray}
|\Psi_n^{SM}(q,\xi,t)|^2\cdot|\Psi_m^{SM}(q,\xi,t)|^2=0 &\textrm{if } n\neq m.
\end{eqnarray} 
The probability $\mathcal{P}_n$ to find  the pointer in the zone $\Delta_n$ of the $\xi-$configuration space is therefore given by 
\begin{eqnarray}
\mathcal{P}_n=\int dq \int_{\Delta_n}d\xi|\Psi^{SM}(q,\xi,t)|^2=|c_n|^2\int dq \int_{\Delta_n}d\xi|\Psi_n^{SM}(q,\xi,t)|^2\nonumber\\
=|c_n|^2\int dq \int d\xi|\Psi_n^{SM}(q,\xi,t)|^2=|c_n|^2\langle \Psi_n^{SM}|\Psi_n^{SM}\rangle=|c_n|^2
\end{eqnarray}
It can be rewritten as
\begin{eqnarray}
\mathcal{P}_n=\langle \Psi_t^{SM}|\hat{\Pi}^{S}_n|\Psi_t^{SM}\rangle=\langle \Psi_0^{SM}|(\hat{U}^{SM})^{-1}\hat{\Pi}^{S}_n\hat{U}^{SM}|\Psi_0^{SM}\rangle
\end{eqnarray} where $\hat{\Pi}^{S}_n=\int_{\Delta_n}d\xi|\xi\rangle\langle\xi|$ is a sum of projectors in the cell $\Delta_n$.
It is equivalent to 
\begin{eqnarray}
\mathcal{P}_n=\langle \psi_0^{S}|\hat{O}_n^S|\psi_0^{S}\rangle
\end{eqnarray}
where $\hat{O}_n^S$ is a POVM defined by 
\begin{eqnarray}
\hat{O}_n^S=\langle \Phi_0^{M}|(\hat{U}^{SM})^{-1}\hat{\Pi}^{S}_n\hat{U}^{SM}|\Phi_0^{M}\rangle
\end{eqnarray} that explicitly reads
\begin{eqnarray}
\hat{O}_n^S=\iint dq_f dq_0|q_f\rangle\langle q_0|\mathcal{M}^{SM}(q_f,q_0)
\end{eqnarray} with
\begin{eqnarray}
\mathcal{M}^{SM}(q_f,q_0)=\int_{\Delta_n}d\xi\iiint dqd\xi_fd\xi_0{\Phi_0^{M}}^\ast(\xi_f)\Phi_0^{M}(\xi_0)K^{SM}(q,\xi;q_0,\xi_0){K^{SM}}^\ast(q,\xi;q_f,\xi_f)
\end{eqnarray} and the propagator $K^{SM}(q,\xi;q_0,\xi_0)=\langle q,\xi|\hat{U}^{SM}|q_0,\xi_0\rangle$. We stress that we have the condition $(\mathcal{M}^{SM}(q_0,q_f))^{\ast}=\mathcal{M}^{SM}(q_f,q_0)$ that implies the self-adjointeness $\hat{O}_n^S=(\hat{O}_n^S)^\dagger$ required in the definition of a POVM. To complete our definition of a dBB POVM we observe that we have $\sum_n \hat{O}_n^S=\hat{I}$, and $\langle \psi_0^{S}|\hat{O}_n^S|\psi_0^{S}\rangle\geq 0$ whatever $|\psi_0^{S}\rangle$.\\
\indent The previous analysis was limited to spinless systems. If we consider systems of particles with spins we replace the wave function $\Psi^{SM}(q,\xi,t)$ by $\Psi_{i^S,j^M}^{SM}(q,\xi,t)$ where $i^S$ and $j^M$ are discrete spin indices for the $S$ and $M$ subsystems. From Eq.~\ref{POVM1} we still have $\mathcal{P}_n=\langle \psi_0^{S}|\hat{O}_n^S|\psi_0^{S}\rangle$ where the POVM reads 
\begin{eqnarray}
\hat{O}_n^S=\sum_{i_f^S,i_0^S}\iint dq_f dq_0|q_f\rangle\langle q_0|\mathcal{M}_{i_f^S,i_0^S}^{SM}(q_f,q_0)
\end{eqnarray} with $\mathcal{M}_{i_f^S,i_0^S}^{SM}(q_f,q_0)=(\mathcal{M}_{i_0^S,i_f^S}^{SM}(q_0,q_f))^{\ast}$ such that
\begin{eqnarray}
\mathcal{M}_{i_f^S,i_0^S}^{SM}(q_f,q_0)=\sum_{i^S,j^M,j_f^M,j_0^M}\int_{\Delta_n}d\xi\iiint dqd\xi_fd\xi_0{\Phi_{0,j_f^M}^{M}}^\ast(\xi_f)\Phi_{0,j_0^M}^{M}(\xi_0)\nonumber\\ K_{i^S,j^M|i_0^S,j_0^M}^{SM}(q,\xi;q_0,\xi_0){K^{SM}}_{i^S,j^M|i_f^S,j_f^M}^\ast(q,\xi;q_f,\xi_f)
\end{eqnarray} and $K_{i^S,j^M|i_0^S,j_0^M}^{SM}(q,\xi;q_0,\xi_0)=\langle q,\xi,i^S,j^M|\hat{U}^{SM}|q_0,\xi_0,i_0^S,j_0^M\rangle$.
\section[\appendixname~\thesection]{Scattering by an absorbing Fabry-Perot detector}
\label{appendix1}
\indent We model the absorbing medium by a set of atoms with individual extinction cross-section $\sigma_{ext}=\frac{4\pi}{k}\textrm{Im}[f_0]$ where $k=mv$ is the momentum of the incident particle and $f_0$ the (complex valued) inelastic scattering amplitude associated with a spherically symmetric wave $\Psi_s\simeq \frac{f_0e^{ikr}}{r}$. In the regime where the density $N$ of absorbing atom is not too high the wave function propagating in the medium obeys the equation 
\begin{eqnarray}
\boldsymbol{\nabla}^2 \Psi+ (k^2+4\pi f_0 N)\Psi=\boldsymbol{\nabla}^2 \Psi+ 2m(E-V_{eff})\Psi=0 
\end{eqnarray} corresponding to a medium having  an effective propagation index $n_{eff}=\sqrt{[1+\frac{4\pi f_0 N}{k^2}]}$, i.e., to a medium characterized by an effective (complex valued) potential $V_{eff}=-\frac{4\pi f_0 N}{2m}$ with $\textrm{Im}[V_{eff}]=-N\frac{k}{2m}\sigma_{ext}<0$. The time-dependent Schr\"odinger evolution in this potential $i\partial_t \Psi_t=[\frac{-\boldsymbol{\nabla}^2}{2m}+V_{eff}]\Psi_t$ leads to the conservation law Eq.~\ref{fluxdis} containing a dissipation term $2\textrm{Im}[V_{eff}]|\Psi|^2$ due to the violation of unitarity in this effective model.\\
\indent We now consider a plane wave incident on such a medium supposed to be confined in a (Fabry-Perot) slab between the parallel surfaces $z=0$ and $z=d$. The incident plane wave reads $\Psi^{(0)}=e^{ik_1 z}e^{ik_{x}x}e^{-i\frac{k^2}{2m}t}$ where  $k_x=k\sin{\theta}$, $k_1=k\cos{\theta}=\sqrt{k^2-k_{x}^2}$ are respectively the  $x$ and $z$ wavevector components, $\theta$ is the incidence angle, and $k$ the wavevector associated with the kinetic energy $\frac{k^2}{2m}$.   In presence of the Fabry-Perot slab the wave function in the region $z<0$ and $z>d$ read respectively:
\begin{eqnarray}
\Psi_{<}=(e^{ik_1 z}+Re^{-ik_1 z})e^{ik_x x}e^{-i\frac{k^2}{2m}t}\nonumber\\
\Psi_{>}=Te^{ik_1 z}e^{ik_x x}e^{-i\frac{k^2}{2m}t}
\end{eqnarray} 
Fresnel's reflection and transmission coefficients $R$, $T$ are given by standard formulas:
\begin{eqnarray}
R=\frac{r}{1-r^2e^{i\delta}}(1-e^{i\delta})\nonumber\\
T=\frac{k_2}{k_1}\frac{te^{i\delta/2}}{1-r^2e^{i\delta}}
\end{eqnarray}
 where $r=\frac{k_1-k_2}{k_1+k_2}$, $t=2\frac{k_1}{k_1+k_2}$ are the single interface Fresnel's coefficients (with $k_2=\sqrt{k_1^2+4\pi f_0 N}$ the $z-$wavevector component in the absorbing medium), and $\delta=2k_2 d$ is a complex phase shift.\\ 
 \indent In the medium, for $0<z<d$, the wave function reads $
\Psi_{\textrm{inside}}=(Ce^{ik_2 z}+De^{-ik_2 z})e^{-ik_x x}e^{-i\frac{k^2}{2m}t}$ with 
\begin{eqnarray}
C=\frac{1}{2}[1+R+\frac{k_1}{k_2}(1-R)]\nonumber\\
D=\frac{1}{2}[1+R-\frac{k_1}{k_2}(1-R)]
\end{eqnarray}
\indent The dBB trajectories can be computed in the different regions using the probability current $\mathbf{J}^\Psi=\textrm{Im}[\Psi^\dagger\boldsymbol{\nabla}\Psi]/m$.
We have for $z<0$ 
\begin{eqnarray}
J^\Psi_z=\frac{k_1}{m}(1-|R|^2),\nonumber\\
J^\Psi_x=\frac{k_x}{m}(1+|R|^2+2|R|\cos{(2k_1z-arg(R))})
\end{eqnarray} 
leading to the trajectory equation $\frac{dz}{dx}=\frac{J^\Psi_z}{J^\Psi_x}$ in the interfering region:
\begin{eqnarray}
\frac{dz}{dx}=\cot{\theta}\frac{1-|R|^2}{1+|R|^2+2|R|\cos{(2k_1z-arg(R))}}
\end{eqnarray} 
The mean trajectory, around which the particle oscillates, obeys the equation $\frac{dz}{dx}=\cot{\theta}\frac{1-|R|^2}{1+|R|^2}$ which has a geometrical interpretation as shown in [XX] and Figure \ref{Fig1}. In the slab for $0<z<d$ we similarly obtain
\begin{eqnarray}
\frac{dz}{dx}=\frac{\frac{k'_2}{kx}(|C|^2e^{-2k"_2z}-|D|^2e^{2k"_2z})+2\frac{k"_2}{kx}|DC|\sin{(\xi)}}{|C|^2e^{-2k"_2z}-|D|^2e^{2k"_2z}+2|DC|\cos{(\xi)}}\nonumber\\
\end{eqnarray}  with  $\xi=2k'_2z+arg(D)-arg(C)$, $k'_2=\textrm{Re}[k_2]$, $k"_2=\textrm{Im}[k_2]$. This defines a very complicated motion [XX]. In the transmitted region $z>d$ we have $\frac{dz}{dx}=\cot{\theta}$ as it should be.\\ 
 \indent From Eq.~\ref{fluxdis} we can calculate the difference between the probability current flows through the surfaces $z=0$ and $z=d$
\begin{eqnarray}
\mathcal{I}_{z=0}-\mathcal{I}_{z=d}=\Sigma\cdot v\cos{\theta}[1-|R|^2-|T|^2]
=\Sigma\cdot N\sigma_{ext}v\int_{z=0}^{z=d}dz|\Psi|^2(x,y,z)
\end{eqnarray} 
 with $\mathcal{I}_{z=0}=\int_\Sigma dxdyJ^\Psi_z(x,y,z=0)$, $\mathcal{I}_{z=d}=\int_\Sigma dxdyJ^\Psi_z(x,y,z=d)$ and $\Sigma$ is the whole lateral surface of the slab. Importantly $\Sigma\cdot N\sigma_{ext}v\int_{z=0}^{z=d}dz|\Psi|^2(x,y,z)$ represents the probability of absorption by the slab per unit time, i.e., it defines the fraction of incident particles  trapped by the detector per unit time or the arrival time probability density $\mathcal{P}^\Psi(\Sigma,\tau)$ (here the situation is time independent).\\
\indent  Two extreme regimes are relevant for the present discussion. First, in the weak coupling regime with a low density $N$ and small cross-section $\sigma_{ext}$ we have a semi transparent medium $r\simeq 0 $ implying $R\simeq 0$ and $T\simeq e^{ik_2 d}$, i.e., $|T|^2\simeq e^{-\frac{N\sigma_{ext}d}{\cos{\theta}}}$. We thus get  
\begin{eqnarray}
\mathcal{P}^\Psi(\Sigma,\tau)\simeq\Sigma\cdot N d \sigma_{ext}v =\Sigma\cdot N d \sigma_{ext}|\mathbf{J}^{\Psi^{(0)}}|
\end{eqnarray} which is proportional to the norm of the initial probability current and doesn't depend on the incidence angle $\theta$. In the second `strong absorption' regime we assume 
  $\textrm{Im}[\delta]\gg 1$ and thus $|T|^2\simeq|t|^4|\frac{k_2}{k_1}|^2e^{-2\textrm{Im}[\delta]}\rightarrow 0$ and $|R|^2\rightarrow |r|^2$.    We thus get the 	arrival-time density of probability:
 \begin{eqnarray}
\mathcal{P}^\Psi(\Sigma,\tau)\simeq\Sigma\cdot v\cos{\theta}[1-|r|^2]=\Sigma\cdot J_z^{\Psi^{(0)}}[1-|r|^2]
\end{eqnarray} Note that in the limit where the medium is strongly absorbing we have $r\rightarrow -1$ and therefore the probability of absorbing a particle tends to vanish aswell.
\section[\appendixname~\thesection]{Perfectly matched layer detectors: general derivations}
\label{appendix2}
\indent We write
\begin{eqnarray}
\Psi^{(abs)}=e^{ik_zf(z)}e^{i\mathbf{k}_{||}\cdot \mathbf{x}_{||}}e^{-i\frac{k^2}{2m}t}.\label{truc}
\end{eqnarray}
We immediately check that we have 
\begin{eqnarray}
\frac{1}{f'(z)}\partial_z(\frac{1}{f'(z)}\partial_z e^{ik_zf(z)})=-k_z^2e^{ik_zf(z)}
\end{eqnarray} or equivalently
\begin{eqnarray}
\partial_z^2\Psi^{(abs)}-\frac{f"(z)}{f'(z)}\partial_z\Psi^{(abs)}+k_z^2(f'(z))^2\Psi^{(abs)}=0.
\end{eqnarray} The first order derivative can be eliminated by using the condition $\frac{1}{f'(z)}\partial_z e^{ik_zf(z)})=ik_ze^{ik_zf(z)}$ and therefore we have 
\begin{eqnarray}
\partial_z^2\Psi^{(abs)}-ik_zf"(z)\Psi^{(abs)}+k_z^2(f'(z))^2\Psi^{(abs)}=0.
\end{eqnarray} 
Writing $k_zf(z)=k_zz+i\int_{-\infty}^zdz'\chi(z')$ ($\chi(z)$ defining the absorption of the system), $f'(z)=1+i\frac{\chi(z)}{k_z}$, $f"(z)=i\frac{\chi'(z)}{k_z}$, $(\partial^2_x+\partial^2_y)
\Psi^{(abs)}=-\mathbf{k}^2_{||}\Psi^{(abs)}$ we thus deduce
\begin{eqnarray}
\boldsymbol{\nabla}^2\Psi^{(abs)}(z,\mathbf{x}_{||},t)+2m(E-V_{eff}(z))\Psi^{(abs)}(z,\mathbf{x}_{||},t)=0\nonumber\\
\end{eqnarray} with the effective complex potential
\begin{eqnarray}
V_{eff}(z)=\frac{\chi^2(z)-\chi'(z)}{2m}-i\chi(z)\frac{k_z}{m}.\label{potential}
\end{eqnarray}
\section[\appendixname~\thesection]{Perfectly matched layer  detectors: a particular model}
\label{appendix3}
\indent In relation with Appendix \ref{appendix2} we now impose
\begin{eqnarray}
\chi(z)= \chi_0\cdot[ \theta(-z)e^{-az^2}+\theta(z)\theta(d-z)+\theta(z-d)e^{-a(z-d)^2}]\label{renorma1}
\end{eqnarray} where $1/\sqrt{a}$ defines a characteristic length over which the potential $V_{eff}$ rises continuously around the two zones $z\simeq 0$ and $d\simeq d$.
With this choice the function $f(z)=z+\frac{i}{k_z}\int_{-\infty}^zdz'\chi(z')$ in Eq.~\ref{truc} reads 
\begin{eqnarray}
f(z)=z+\frac{i}{2k_z}\chi_0\sqrt{(\frac{\pi}{a})}[1+erf(\sqrt{a} z)] & \textrm{if } z\leq 0 \nonumber\\
f(z)=z+\frac{i}{k_z}\chi_0[z+\frac{1}{2}\sqrt{(\frac{\pi}{a})}] & \textrm{if } 0 \leq z\leq d \nonumber\\
f(z)=z+\frac{i}{k_z}\chi_0[d+\frac{1}{2}\sqrt{(\frac{\pi}{a})}(1+erf(\sqrt{a} (z-d)))] & \textrm{if } d\leq z.
\end{eqnarray}
where $erf(x)=\frac{2}{\sqrt{\pi}}\int_0^xdze^{-z^2}$.
From Eq.~\ref{potential} we deduce $V_{eff}$ with $\chi'(z)= -2\chi_0a\cdot[z\theta(-z)e^{-az^2}+(z-d)\theta(z-d)e^{-a(z-d)^2}]$. As shown in Fig.\ref{Fig2} the potential is a continuous function of $z$ (with slope dicontinuities at $z=0$ and $z=d$ arising from the second order derivative $\chi''(z)$). \\
\indent In analogy with Eq.~\ref{detexB} we define the probability $\mathcal{P}_{\textrm{detec.}}^\Psi(\Sigma,t)$
\begin{eqnarray}
\mathcal{P}_{\textrm{detec.}}^\Psi(\Sigma,t)=2\Sigma\frac{k}{m}\int_{-\infty}^{+\infty}dz\chi(z)e^{-2\int_{-\infty}^zdz\chi(z)}  \label{detexE}
\end{eqnarray}which reads
\begin{eqnarray}
\mathcal{P}_{\textrm{detec.}}^\Psi(\Sigma,t)=\Sigma\frac{k}{m}[e^{-\xi}(1-e^{-2\chi_0d})+\xi F(\xi)+e^{-2\chi_0d}\xi G(\xi)]  \label{probaFi}
\end{eqnarray} with $\xi=\chi_0\sqrt{(\frac{\pi}{a})}$, $F(\xi)=\frac{2}{\sqrt{\pi}}\int_{-\infty}^0dze^{-z^2}e^{\xi(1+erf(z))}$, and $G(\xi)=\frac{2}{\sqrt{\pi}}\int_0^{+\infty}dze^{-z^2}e^{\xi(1+erf(z))}$.  In the limit $\xi\rightarrow 0$  (i.e. $a\rightarrow +\infty$) we have $F(0)=G(0)=1$ and we recover the result Eq.~\ref{detexC}.\\
\indent The present analysis for a detector adapted to a plane wave $\propto e^{+ik_z z}$ with $k_z>0$ can be used to define the absorbing medium  corresponding to an incident plane wave propagating in the opposite direction, i.e., $\propto e^{-ik_z z}$. For this we write the previous solution  Eq.~\ref{truc} as $\Psi^{(abs)}=e^{ik_z z}e^{-F(z)}e^{i\mathbf{k}_{||}\cdot \mathbf{x}_{||}}e^{-i\frac{k^2}{2m}t}$ and we define the counterpropagating wave   as the one obtained under the transformation $z\rightarrow d-z$. We write the new wave:  
\begin{eqnarray}
\tilde{\Psi}^{(abs)}=e^{-ik_z z}e^{-F(d-z)}e^{i\mathbf{k}_{||}\cdot \mathbf{x}_{||}}e^{-i\frac{k^2}{2m}t}=e^{-ik_z \tilde{f}(z)}e^{i\mathbf{k}_{||}\cdot \mathbf{x}_{||}}e^{-i\frac{k^2}{2m}t}
\end{eqnarray} where we omitted a phase constant.
We have the transformation $F(Z)=\int_{-\infty}^Zdz\chi(z)\rightarrow F(d-Z)=\int_Z^{+\infty}dz\chi(d-z)$ and therefore $\tilde{f}(Z)=Z-i\frac{1}{k_z}\int_Z^{+\infty}dz\chi(d-z)$, $\tilde{f}'(Z)=1+i\frac{1}{k_z}\chi(d-Z)$, $\tilde{f}"(Z)=-i\frac{1}{k_z}\frac{d}{dz}\chi(z)|_{z=d-Z}$.
Finally we deduce
\begin{eqnarray}
\partial_z^2\tilde{\Psi}^{(abs)}+ik_z\tilde{f}"(z)\tilde{\Psi}^{(abs)}+k_z^2(\tilde{f}'(z))^2\tilde{\Psi}^{(abs)}=0.
\end{eqnarray} and therefore
\begin{eqnarray}
\boldsymbol{\nabla}^2\tilde{\Psi}^{(abs)}(z,\mathbf{x}_{||},t)+2m(E-\tilde{V}_{eff}(z))\tilde{\Psi}^{(abs)}(z,\mathbf{x}_{||},t)=0\nonumber\\
\end{eqnarray} with the new effective complex potential adapted to the counterpropagative wave:
\begin{eqnarray}
\tilde{V}_{eff}(Z)=\frac{\chi^2(d-Z)-\frac{d}{dz}\chi(z)|_{z=d-Z}}{2m}-i\chi(d-Z)\frac{k_z}{m}.\label{potentialnew}
\end{eqnarray} 
With the example of Eq.~\ref{renorma1} we  have $\chi(d-z)=\chi(z)$,   and $\frac{d}{dz}\chi(z)|_{z=d-Z}=-\chi'(Z)$ and therefore
\begin{eqnarray}
\tilde{V}_{eff}(Z)=\frac{\chi^2(Z)+\chi'(Z)}{2m}-i\chi(Z)\frac{k_z}{m}.\label{potentialnewb} 
\end{eqnarray} This can be compared  with Eq.~\ref{potential} for the choice Eq.~\ref{renorma1}: The two effective potentials differ by  the sign in front of $\chi'(z)$.
\section[\appendixname~\thesection]{Backflow with two plane waves}
\label{appendix4}
\indent From \begin{eqnarray}
\Psi^{(0)}(\mathbf{x},t)=(e^{i\mathbf{k}_1\cdot\mathbf{x}}-\frac{1}{2}(1+\frac{k_{1z}}{k_{2z}}) e^{i\mathbf{k}_2\cdot\mathbf{x}})e^{-iEt}
\end{eqnarray} obtained with $\alpha=\alpha_{min}=-\frac{1}{2}(1+\frac{k_{1z}}{k_{2z}})$ in Eq.~\ref{backflow} we deduce the probability current $z-$component at the point $\mathbf{x}_0=0$:
 \begin{eqnarray}
J_z^{\Psi^{(0)}}(\mathbf{x}_0=0)=\frac{k_{2z}}{m}[|\alpha|^2-(1+\frac{k_{1z}}{k_{2z}})|\alpha|+\frac{k_{1z}}{k_{2z}}]=-\frac{k_{2z}}{4m}(1-\frac{k_{1z}}{k_{2z}})^2<0.
\end{eqnarray}
Similarly we have $|\Psi^{(0)}(\mathbf{x}_0=0)|^2=(1-|\alpha|)^2=\frac{1}{4}(1-\frac{k_{1z}}{k_{2z}})^2$. 
This allows us to define an effective ($z-$component) wavevector: 
\begin{eqnarray}
k_{eff,z}(\mathbf{x}_0=0)=m\frac{J_z^{\Psi^{(0)}}(\mathbf{x}_0=0)}{|\Psi^{(0)}(\mathbf{x}_0=0)|^2}=-k_{2z}.
\end{eqnarray}We can easily deduce the other components of  $\mathbf{k}_{eff}(\mathbf{x}_0=0)$. In particular the quantum potential reads
\begin{eqnarray}
Q^{\Psi^{(0)}}(\mathbf{x}_0=0)=\frac{-\boldsymbol{\nabla}^2|\Psi^{(0)}|}{2m|\Psi^{(0)}|}|_{\mathbf{x}_0=0}=\frac{(\mathbf{k}_1-\mathbf{k}_2)^2}{m}\frac{(1+\frac{k_{1z}}{k_{2z}})}{(1-\frac{k_{1z}}{k_{2z}})^2}
\end{eqnarray}
\section[\appendixname~\thesection]{The full arrival time distribution with non efficient detectors}
\label{appendix5}
We start with Eq.~\ref{currentspin} and consider the full signal  $\mathcal{P}_{\textrm{detec., full signal}}^\Psi(\Sigma,t)\simeq \eta\int d^\Sigma_{\mathbf{x}}J_z^{\Psi_{\hat{\mathbf{s}}}}(\mathbf{x},t)$ which reads 
\begin{eqnarray}
\mathcal{P}_{\textrm{detec.,full signal }}^\Psi(\Sigma,t)
=\eta\int_0{R} d\rho\rho \oint d\varphi(|\Phi(\rho,z=L,t)|^2\frac{\partial_zS(\rho,z=L,t)}{m}\nonumber\\+\frac{\hat{\mathbf{s}}\cdot\hat{\boldsymbol{\varphi}}}{2m}\partial_\rho|\Phi(\rho,z=L,t)|^2).
\label{currentspindetec}\end{eqnarray}
Since we are working in the weak coupling regime the current $J_z^{\Psi_{\hat{\mathbf{s}}}}(\mathbf{x},t)$ can be negative an this is associated with back flow. The contributions of back flow is negative in Eq.~\ref{currentspindetec}. However it is not difficult to see that the  second term of the integral associated with the spin-magnetic current vanishes. This is trivially so for the longitudinal case where $\hat{\mathbf{s}}=\pm \hat{\mathbf{z}}$. For the transverse cases it is sufficient to consider the case  $\hat{\mathbf{s}}=+ \hat{\mathbf{x}}$ the other cases beeing equivalent due to rotational invariance of the problem. If $\hat{\mathbf{s}}=+ \hat{\mathbf{x}}$
 we have 
 \begin{eqnarray}
\oint d\varphi\frac{\hat{\mathbf{s}}\cdot\hat{\boldsymbol{\varphi}}}{2m}\partial_\rho|\Phi(\rho,z=L,t)|^2=\oint d\varphi\frac{\cos{(\varphi)}}{2m}\partial_\rho|\Phi(\rho,z=L,t)|^2=0
\label{currentspindetec}\end{eqnarray} as required . Therefore we have 
\begin{eqnarray}
\mathcal{P}_{\textrm{detec.,full signal }}^\Psi(\Sigma,t)
=\eta\int_0{R} d\rho\rho \oint d\varphi|\Phi(\rho,z=L,t)|^2\frac{\partial_zS(\rho,z=L,t)}{m}.
\label{currentspindetec}\end{eqnarray} which is spin independent and considers only the convective current.  We have recovered DD result \cite{DasDurr2019} (see Eq.~\ref{detexDD}):
 \begin{eqnarray}
\mathcal{P}_{\textrm{detec.,full signal }}^\Psi(\Sigma,\tau)
=\eta \frac{4L^3}{\lambda_0 \sqrt{\pi}}\frac{\tau e^{-\frac{L^2}{1+\tau^2}}}{(1+\tau^2)^{5/2}}.
\label{detexDDnew} 
\end{eqnarray}



\end{document}